\documentclass[12pt]{article}
\usepackage{amsmath}
\usepackage{graphicx}
\usepackage{enumerate}
\usepackage{natbib}
\usepackage{amssymb}
\usepackage{url} 
\usepackage{algorithm}
\usepackage{algpseudocode}
\usepackage{setspace}
\usepackage{mathdots}
\usepackage{textcomp}
\usepackage{blkarray, bigstrut}
\usepackage{amsthm}
\usepackage{mathtools}
\usepackage{xcolor}
\usepackage{booktabs}
\usepackage{multirow}
\usepackage{graphicx}
\usepackage[normalem]{ulem}
\useunder{\uline}{\ul}{}
\usepackage{authblk}
\usepackage{etoolbox}
\usepackage{setspace}

\newcommand{\blind}{0}

\expandafter\def\expandafter\normalsize\expandafter{%
    \normalsize%
    \setlength\abovedisplayskip{0pt}%
    \setlength\belowdisplayskip{4pt}%
    \setlength\abovedisplayshortskip{-8pt}%
    \setlength\belowdisplayshortskip{2pt}%
}

\begin{document}

\bibliographystyle{agsm}

\def\spacingset#1{\renewcommand{\baselinestretch}%
{#1}\small\normalsize} \spacingset{1}



\title{Heterogeneity-Aware Regression with Nonparametric Estimation and Structured Selection for Hospital Readmission Prediction}

\author[a]{Wei Wang}
\author[b]{Angela Bailey}
\author[b]{Christopher Tignanelli}
\author[a,*]{Jared D. Huling} 

\affil[a]{Division of Biostatistics \& Health Data Science, University of Minnesota Twin Cities, Minneapolis, MN, 55414}
\affil[b]{Department of Surgery, University of Minnesota Twin Cities, Minneapolis, MN, 55455}

\date{} 
  \maketitle


\if0\blind
{
  \bigskip
  \bigskip
  \bigskip
  \begin{center}
    \vspace{-1em}
{
\noindent
$^*$Corresponding author(\texttt{huling@umn.edu})
}
\end{center}
  \medskip
} \fi

\bigskip
\begin{abstract}

Readmission prediction is a critical but challenging clinical task, as the inherent relationship between high-dimensional covariates and readmission is complex and heterogeneous. Despite this complexity, models should be interpretable to aid clinicians in understanding an individual's risk prediction. Readmissions are often heterogeneous, as individuals hospitalized for different reasons, particularly across distinct clinical diagnosis groups, exhibit materially different subsequent risks of readmission. To enable flexible yet interpretable modeling that accounts for patient heterogeneity, we propose a novel hierarchical-group structure kernel that uses sparsity-inducing kernel summation for variable selection. Specifically, we design group-specific kernels that vary across clinical groups, with the degree of variation governed by the underlying heterogeneity in readmission risk; when heterogeneity is minimal, the group-specific kernels naturally align, approaching a shared structure across groups. Additionally, by allowing variable importance to adapt across interactions, our approach enables more precise characterization of higher-order effects, improving upon existing methods that capture nonlinear and higher-order interactions via functional ANOVA. Extensive simulations and a hematologic readmission dataset (n=18,096) demonstrate superior performance across subgroups of patients (AUROC, PRAUC) over the lasso and XGBoost. Additionally, our model provides interpretable insights into variable importance and group heterogeneity.

\end{abstract}

\noindent%
\vfill

\newpage
\spacingset{1.9} 
\section{Introduction}

\label{sec:intro}

A readmission event occurs when a patient is released or discharged from the hospital and presents back to the hospital within 30 days. Such events are a key quality metric for inpatient hospital care and readmissions cost the U.S. healthcare system over \$26 billion annually \citep{lahijanian_care_2021}. Reducing the 30-day readmission rate across a health system has thus been a high priority for hospitals given its negative impact on care for patients and health systems financially \citep{dhaliwal_reducing_2025, hcup2025}. Despite significant efforts by hospitals to reduce readmission rates, they often remain high among many health systems, indicating significant gaps in the quality of care delivered.

Identifying patients with high readmission risk is thus crucial to reducing preventable readmissions and associated costs. Predictive modeling plays a central role in this process, yet the poor predictive performance of most readmission risk models and their often black-box nature prevents their effective use in decision-making for patients. The inherent noise in the processes that result in patients being readmitted and gaps in information available in the training of risk models partially account for this poor performance, but limitations in the modeling strategies used are another major factor.
According a systematic review on readmission prediction modeling \citep{teo_current_2023}, most approaches and studies (74.5\%) used regression models either as the main method or a baseline comparison method. More flexible machine learning (ML) methods are also used, among which tree-based models, such as fandom forests (RF) and gradient-boosting (often XGBoost), are most common (44\% of all studies and 75\% of studies using ML techniques). Despite frequent utilization, readmission risk models tend to have suboptimal predictive performance. More flexible ensemble tree-based models, while often achieving superior predictive accuracy when validated internally, suffer from reduced interpretability and are prone to overfitting, leading to suboptimal performance when implemented in the real world prospectively. Both interpretability and robustness to unseen data are crucial for readmission prediction, as understanding underlying risk factors informs intervention design, and reliable generalization ensures effectiveness in practice.

\begin{figure}[htbp]
    \centering
    \includegraphics[width=1\textwidth]{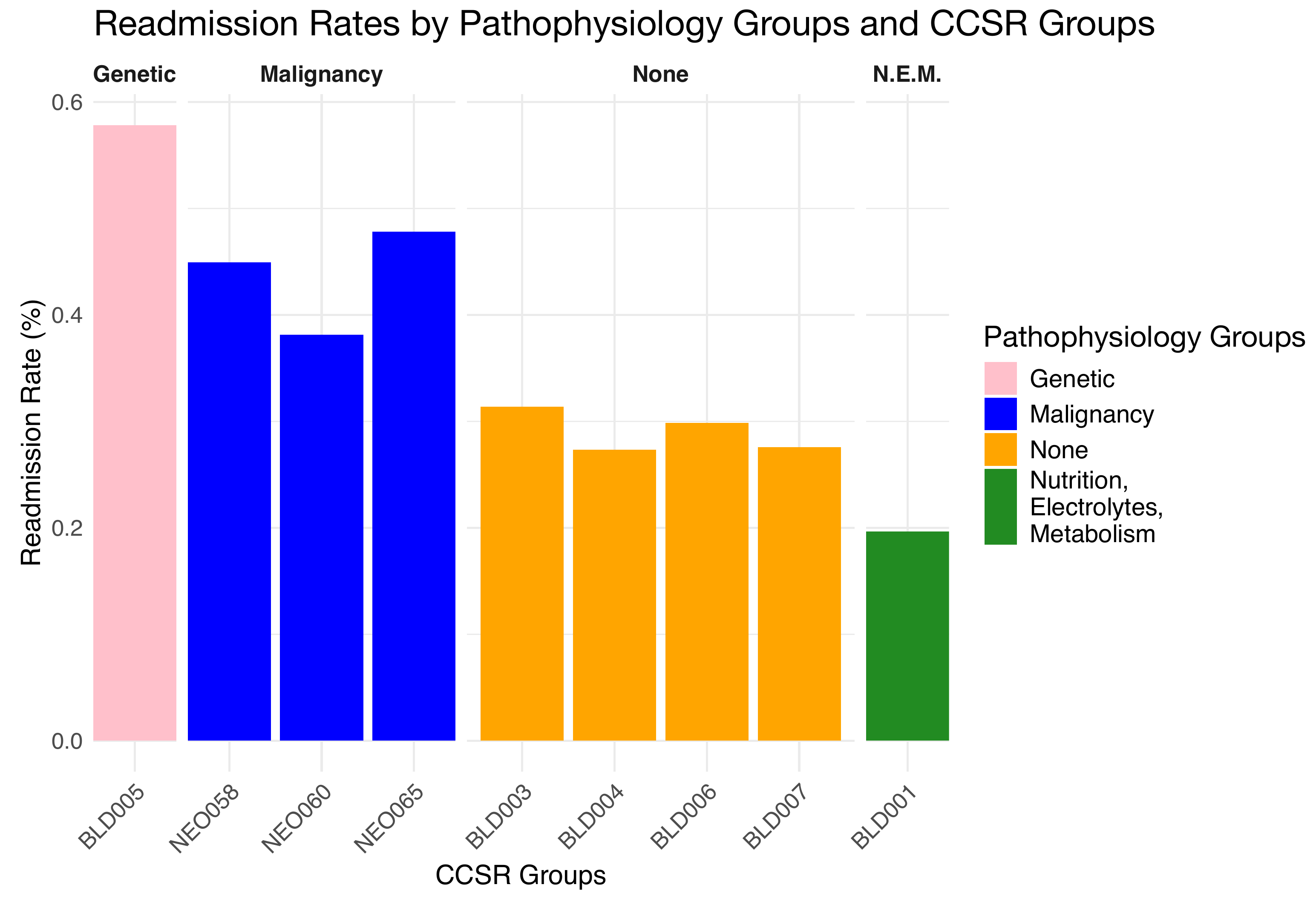}
    \caption{Readmission rates by pathophysiology (level-1) groups and CCSR (level-2) groups for the hematologic population. For prospective relevance, only subgroups with more than 100 samples in the most recent year (2023) are included. Heterogeneity in readmission is observed across groups at both levels.}
    \label{fig:hem_readmit_by_group}
\end{figure}

Another key challenge to the robustness and generalizability of readmission risk prediction modeling is the heterogeneity of patient data from electronic health records (EHRs). Risk models are often built and employed on entire populations in a health system. However, those admitted for different types of health conditions often have markedly different risk profiles. For example, the association of blood pressure with risk may be different for someone admitted for cardiac disorders compared with someone admitted for leg fracture. Ignoring this heterogeneity in risk may result in suboptimal performance and diminish the utility of risk models, since the interpretation of a model developed on a heterogeneous population may be obscured by the conflation of distinct risk profiles. To characterize this heterogeneity, we utilize pre-categorized ICD-10 codes (Centers for Medicare and Medicaid Services based billing codes) by disease system, known as the Clinical Classifications Software Refined (CCSR) \citep{hcup2025}. CCSR groups patients with similar primary diagnoses. The CCSR groups are further aggregated into related pathophysiological groups by clinical subject matter experts. Figure \ref{fig:hem_readmit_by_group} illustrates the heterogeneity in readmission rates across the pathophysiological (level-1) and CCSR (level-2) groups (details in Table \ref{tab:hem_group}) of the hematologic population in a large academic health system. Even within this population, there is substantial observed heterogeneity, motivating the development of methods that can adapt to group-wise heterogeneity of risk.

\begin{table}[]
\caption{Pathophysiology (level-1) groups and CCSR (level-2) groups for the hematologic population.}
\label{tab:hem_group}
\resizebox{\columnwidth}{!}{%
\begin{tabular}{@{}lll@{}}
\toprule
pathophysiology groups (level-1 groups) & CCSR categories (level-2 groups)  & CCSR categories description \\ \midrule
Genetic                             & BLD005 & Sickle cell trait/anemia                               \\ \midrule
Malignancy                          & NEO057 & Hodgkin lymphoma                                       \\
                                    & NEO058 & Non-Hodgkin lymphoma                                   \\
                                    & NEO059 & Leukemia - acute lymphoblastic leukemia (ALL)          \\
                                    & NEO060 & Leukemia - acute myeloid leukemia (AML)                \\
                                    & NEO061 & Leukemia - chronic lymphocytic leukemia (CLL)          \\
                                    & NEO062 & Leukemia - chronic myeloid leukemia (CML)              \\
                                    & NEO063 & Leukemia - hairy cell                                  \\
                                    & NEO064 & Leukemia - all other types                             \\
                                    & NEO065 & Multiple myeloma                                       \\
                                    & NEO068 & Myelodysplastic syndrome (MDS)                         \\ \midrule
None                                & BLD003 & Aplastic anemia                                        \\
                                    & BLD004 & Acute posthemorrhagic anemia                           \\
                                    & BLD006 & Coagulation and hemorrhagic disorders                  \\
                                    & BLD007 & Diseases of white blood cells                          \\
                                    & BLD010 & Other specified and unspecified hematologic conditions \\ \midrule
Nutrition, Electrolytes, Metabolism & BLD001 & Nutritional anemia                                     \\
                                    & BLD002 & Hemolytic anemia                                       \\ \bottomrule
\end{tabular}%
}
\end{table}

Although hierarchical functional ANOVA \citep{stone_use_1994, huang_projection_1998} has been rarely employed in previous readmission risk prediction studies, it is a promising candidate due to its ability to capture complex structures while providing interpretability through hierarchical decomposition of the regression function's variance. Functional ANOVA methods were developed in nonparametric setting via a smoothing spline ANOVA (SS-ANOVA) approach \citep{gu_smoothing_1993}, which implements the decomposition in a reproducing kernel Hilbert space (RKHS) with smoothness penalties. Extensions have been made to deal with high dimensionality. For example, COSSO \citep{lin_component_2006} applies an L1 penalty on RKHS norms to shrink irrelevant components; \citet{reich_variable_2009} proposes a Bayesian model based on SS-ANOVA for variable selection; and SKIM-FA \citep{agrawal_skim-fa_2023} utilizes sparsity-inducing kernel structures characterized by interpretable variable importance to enable variable selection and ensure interactions are included only when relevant main effects are selected. However, such approaches do not address the forms of population heterogeneity often present in the readmission setting.

We propose a method built upon a functional ANOVA type model that incorporates both variable selection and population heterogeneity in a structured manner. Our approach, which we name Heterogeneity-Aware Regression with Nonparametric Estimation and Structured Selection (HARNESS), targets the challenges mentioned above through: (1) the incorporation of nonlinear and higher-order interactions through functional ANOVA, represented by an interpretable kernel structure to better capture the complex underlying structure of the EHR and patient heterogeneity; (2) selection of variables and evaluation of the order-specific variable importance through sparsity-inducing kernel summation, ensuring a more thorough understanding of risk factors for readmission; (3) allowance of the kernel function to vary across the two-level hierarchical groupings to account for heterogeneity at different pathophysiology and CCSR groups. Additionally, we make HARNESS scalable to large-scale EHR data through divide and recombine (D\&R) techniques and mini-batch gradient descent.  Furthermore, compared to ML models, HARNESS provides greater interpretability in terms of quantifying the importance of variables at different interaction orders and the heterogeneity effects for different clinical subgroups in the context of readmission prediction, offering insights into preventative strategies.

\section{Review of Existing Methods}
\label{sec:meth}

\label{sec:headings}
 
This section introduces several important concepts and methods that we will be building on for our proposed methods. Suppose we observe data $D = \{(\boldsymbol{X}_i, Y_i), i = 1, \dots, n\}$, where $\boldsymbol{X}_i \in \mathcal{X}$, a compact set of $\mathcal{R}^p, Y_i \in \{-1, 1\}. (\boldsymbol{X}_i, Y_i)$ are assumed to be i.i.d. with  
$Y_i | \boldsymbol{X}_i \sim \text{Bernoulli}(\pi(\boldsymbol{x}_i))$, and $p(Y_i = 1 | \boldsymbol{X}_i) = \pi(\boldsymbol{x}_i) = \sigma(f(\boldsymbol{x}_i))$,
where $\sigma(\cdot):\mathcal{R} \rightarrow (0,1)$ is a sigmoid function such as $\exp(\cdot) / (1+\exp(\cdot))$.

Suppose that $\boldsymbol{f}$ lies in some RKHS $\mathcal{H}$. In order to perform variable selection, the following loss function with regularization is considered:
\begin{equation}
    \begin{aligned}\label{eq:obj}
    \boldsymbol{\hat{f}} = \mathop{\arg \min}\limits_{\boldsymbol{f} \in \mathcal{H}} \mathcal{L}(\boldsymbol{y}, \boldsymbol{f}) + J(\boldsymbol{f}) = \mathop{\arg \min}\limits_{\boldsymbol{f} \in \mathcal{H}} - \sum_{i=1}^n \log(\sigma(y_i f_i)) + \lambda ||\boldsymbol{f}||^2_{\mathcal{H}},
    \end{aligned}
\end{equation}
where $-\sum_{i=1}^n \log(\sigma(y_i f_i))$ is the negative log likelihood; $||\boldsymbol{f}||_{\mathcal{H}} = \langle \boldsymbol{f}, \boldsymbol{f} \rangle_{\mathcal{H}}$ is RKHS norm of $\boldsymbol{f}$ to induce smoothness in the estimated function; and $\lambda$ is the hyperparameter controlling the weight for the smoothness penalty term.
 
The form of $\mathcal{H}$ dictates the structure of the functions and thus, to incorporate a desired structure, one can consider the following steps: (1) decompose the $\boldsymbol{f}$ as the summation of functions \{$\boldsymbol{f}_V: V \subseteq [p]$ \} that depend on only part of the covariates $\boldsymbol{X}_V$; (2) define the RKHS $\mathcal{H}_V$ that each $\boldsymbol{f}_V$ lies in to be mutually orthogonal resulting in identifiable decomposition, forming the functional ANOVA representation \citep{stone_use_1994} with interpretable lower-order effects in high-dimensional setting; (3) define the $\mathcal{H}$ in terms of $\mathcal{H}_V$.

In the setting without group heterogeneity, $\boldsymbol{f}$ is decomposed as
\begin{equation}\label{functionANOVA}
    \boldsymbol{f} = \boldsymbol{f}_{\emptyset} + \sum_{j= 1}^p \boldsymbol{f}_{\{j\}}(\boldsymbol{x}_j) + \sum_{j<k}^p \boldsymbol{f}_{\{j, k\}}(\boldsymbol{x}_j, \boldsymbol{x}_k) + \dots ,
\end{equation}
where $\boldsymbol{f}_{\emptyset}$ represents the intercept; $\boldsymbol{f}_{\{j\}}$ represents the main effect of $\boldsymbol{X}_j$; $\boldsymbol{f}_{\{j, k\}}$ represents the two-way effect of $\boldsymbol{X}_j, \boldsymbol{X}_k$; and so on. In order to make the decomposition identifiable, {$\mathcal{H}_V$ can be constructed for $\boldsymbol{f}_V$ as
\begin{equation*}
    \mathcal{H}_V = \{\boldsymbol{f}_V: \forall A \subsetneq V,  \forall f_A, \langle\boldsymbol{f}_V, \boldsymbol{f}_A\rangle = 0 \},
\end{equation*}
where $\boldsymbol{f}_V: \mathcal{X}_v \rightarrow \mathcal{R}$ are all square-integrable functions of $\boldsymbol{X}_V$ with respect to a probability measure $\mu$ such that $\int_{\mathcal{X}_V} |\boldsymbol{f}(\boldsymbol{x}_V)|^2 \,d\mu_V{(\boldsymbol{x}_V)} < \infty$, for which the probability measure $\mu_V$ can be chosen as (a multiple of) the Lebesgue measure for interpretable results \citep{gu_smoothing_1993}. This construction means that the $\mathcal{H}_V$ includes only $\boldsymbol{f}_V$ whose variance is not explained by some lower-order effects of $\boldsymbol{X}_V$. Equation \eqref{functionANOVA} is an orthogonal decomposition of $\boldsymbol{f}$, which is a functional ANOVA decomposition.

We then consider constructing $\mathcal{H}$ from $\mathcal{H}_V$. For each RKHS $\mathcal{H}_V$, there exists a unique reproducing kernel $K_V$ such that $K_V(\boldsymbol{x}, \tilde{\boldsymbol{x}}) = \langle \phi_V(\boldsymbol{x}), \phi_V(\tilde{\boldsymbol{x}}) \rangle$, where the feature mapping $\phi_V: \mathcal{X} \rightarrow \mathcal{H}_V$ forms the basis of $\mathcal{H}_V$. In order to induce variable selection, we consider to construct the reproducing kernel $K$ for $\mathcal{H}$, according to \citet{agrawal_skim-fa_2023}, as
\begin{equation*}
    K(\boldsymbol{x}, \tilde{\boldsymbol{x}}) = \sum_V \theta_V K_V(\boldsymbol{x}, \tilde{\boldsymbol{x}}),
\end{equation*}
where $\theta_V \in [0, \infty)$ can be intuitively interpreted  as a scaling factor that scales the relative contribution of each $\boldsymbol{f}_V$ to the overall $\boldsymbol{f}$. When $\theta_V = 0$, the effect of $\boldsymbol{X}_V$ is removed. To explain why this is the case, for kernel $K'_V = \theta_V K_V$, the corresponding feature map $\phi_V(\cdot)$, which maps a $|V|$-dimensional input matrix into $B_V$-dimensional feature space, changes accordingly as $\phi'_V(\boldsymbol{x}_V) = \sqrt{\theta_V} \cdot \phi_V(\boldsymbol{x}_V)$. From the weight-space view, $\boldsymbol{f} = \sum_V \phi'_V(\boldsymbol{x}_V) \boldsymbol{w}'_V$, so the larger $\theta_V$ indicates that the corresponding feature map $\phi_V'(\boldsymbol{x}_V)$ will be more dominant among the feature maps of $\boldsymbol{X}$, which is the concatenation of all $\phi'_V(\boldsymbol{x}_V)$.

Now the reproducing kernel $K$ and the corresponding RKHS $\mathcal{H}$ that $\boldsymbol{f}$ lies in are constructed. However, this expression requires combinatorial summation to evaluate which can be slow in practice. To further reduce the time complexity, as in \citet{agrawal_skim-fa_2023}, $\theta_V$ is reparametrized as
\begin{align*}
    \theta_V = \eta^2_{|V|} \prod_{j \in V} \kappa_j, \; \text{ and thus } \;
    K(\boldsymbol{x}, \tilde{\boldsymbol{x}}) = \sum_{V} \eta^2_{|V|} \prod_{j \in V} \kappa_j K_V(\boldsymbol{x}, \tilde{\boldsymbol{x}}),
\end{align*}
where $\kappa_j$ quantifies covariate ``importance", and $\kappa_j = 0$ if $\boldsymbol{f}$ does not depend on covariate $\boldsymbol{X}_j$. Due to the term $\prod_{j \in V} \kappa_j$, $\boldsymbol{f}_V$ is not selected as long as there is $j \in V$ such that $\kappa_j = 0$. Intuitively this can be thought as if a particular covariate is not selected, then all the higher-order effects involving it will not be selected accordingly. The terms $\eta_{|V|}^2$ quantifies the overall strength of the $|V|$-way interactions. $\boldsymbol{\kappa}$ and $\boldsymbol{\eta}$ are treated as learnable parameters that will be optimized during training. The illustration for the effect of variable importance $\boldsymbol{\kappa}$ is in Section 1 of the supplementary materials.

\section{New Methodology}

Albeit achieving relative balance between interpretability and flexibility, the kernel defined in Section \ref{sec:meth} utilizes a shared $\kappa_j$ for each $\boldsymbol{X}_j$ across all orders of interactions and subgroups, ignoring potential inconsistencies between effect orders and heterogeneity across subgroups.

We propose a kernel with two key features designed to address the above limitations: (1) the introduction of order-specific variable importance; (2) the incorporation of the the group-wise heterogeneity through the heterogeneity-aware functional decomposition and the corresponding kernel construction. These two modifications (1) allow more flexible structure for the measure of variable importance, which encourages more specific interpretations of the effects on different orders and thus more tailored advice on potential interventions to be considered in the future; (2) account for hierarchical subgroup structure naturally characterize the hematologic population, which not only helps with improving  predictive performance but also interpreting heterogeneity across clinical subgroups. We propose a regression framework based on this kernel, which we call Heterogeneity-Aware Regression with Nonparametric Estimation and Structured Selection (HARNESS).

In the remainder of this section we first introduce techniques for selecting variables and downweighting the influence of unimportant covariates on the estimated regression function. We then introduce our proposed modification to the kernel structure to incorporate hierarchical group-structured heterogeneity in a manner that allows covariate selection globally and locally at the group level.

Suppose we observe data $D' = \{(\boldsymbol{X}_i, \boldsymbol{Z}_i, Y_i), i = 1, \dots, n\}$, where $Y_i \in \{-1, 1\}$ is a binary outcome, $\boldsymbol{X}_i \in \mathcal{X}$ is a vector of covariates of dimension $p$, $\boldsymbol{Z}_i \in \mathcal{Z} = \{0, 1\}^{p_g}$ is an indicator vector that indicates what combination of hierarchically-structured groups an individual is in. $\boldsymbol{Z}_i$ is of dimension $p_g$ for a $Q_g$-level hierarchical group structure and forms the group design matrix $\boldsymbol{Z}$. To explain how $\boldsymbol{Z}$ is constructed to encode the hierarchical group information precisely, first consider denoting the group information at level $k \in \{1, \cdots, Q_g\}$ by $\boldsymbol{Z}_k'$, where $(\boldsymbol{Z}_k')_{i \cdot} \in \{0, 1\}^{p_{g, k}'}$ with $p_{g, k}'$ indicating the total number of possible groups that individuals can belong to at level $k$. To incorporate interactions among group structures across hierarchical levels (higher-level groups and their nested subgroups), we consider to encode the group information from level-1 up to level-$k$ through $\boldsymbol{Z}_k := \boldsymbol{Z}_1' \otimes \cdots \otimes \boldsymbol{Z}_k'$ for $2\le k \le Q_g$ and $\boldsymbol{Z}_1 = \boldsymbol{Z}_1'$. Finally, $\boldsymbol{Z} := \begin{bmatrix} \boldsymbol{Z}_1 & \boldsymbol{Z}_2 & \cdots & \boldsymbol{Z}_{Q_g} \end{bmatrix}$ of dimension $p_g = \sum_{k = 1}^{Q_g} \prod_{l = 1}^k p_{g, l}'$ accordingly.

For the following, we set $Q_g = 2$ in consistency with the hematologic data structure. As an illustrative example: sample size $n = 4$ with half of them in level-1 group 1 and the other half in level-1 group 2; and individuals in each level-1 group belong to two different level-2 groups respectively. The group design matrix in this case is
\begin{equation*}
  \boldsymbol{Z}= \left[ \boldsymbol{Z}_1 \,\middle|\, \boldsymbol{Z}_1' \otimes \boldsymbol{Z}_2' \right] = 
  \begin{blockarray}{*{6}{c} l}
    \begin{block}{*{6}{>{$\footnotesize}c<{$}} l}
      \text{group 1} & \text{group 2} & \text{group 1-1} & \text{group 1-2} & \text{group 2-1} & \text{group 2-2} & \\
    \end{block}
    \begin{block}{[cc!{\quad\vrule width 0.8pt}cccc]l}
      1 & 0 & 1 & 0 & 0 & 0 & \\
      1 & 0 & 0 & 1 & 0 & 0 & \\
      0 & 1 & 0 & 0 & 1 & 0 & \\
      0 & 1 & 0 & 0 & 0 & 1 &. \\
    \end{block}
  \end{blockarray}
\end{equation*}

We assume that $(\boldsymbol{X}_i, \boldsymbol{Z}_i, Y_i)$ are i.i.d. and further that the outcome's conditional distribution follows $Y_i | \boldsymbol{X}_i, \boldsymbol{Z}_i \sim \text{Bernoulli}(\pi(\boldsymbol{x}_i, \boldsymbol{z}_i))$ with $p(Y_i = 1 | \boldsymbol{X}_i, \boldsymbol{Z}_i) = \pi(\boldsymbol{x}_i, \boldsymbol{z}_i) = \sigma(f(\boldsymbol{x}_i, \boldsymbol{z}_i))$ where $\boldsymbol{f}$ is an unspecified function of covariates and group membership. 
To account for the group heterogeneity, we consider the following heterogeneity-aware decomposition of $\boldsymbol{f}: (\mathcal{X}, \mathcal{Z}) \rightarrow \mathcal{R}$ as
\begin{align*}
    \boldsymbol{f}(\boldsymbol{x, z}) &= \underbrace{\boldsymbol{f}^{\emptyset}(\boldsymbol{x})}_{\text{global effects}} + \underbrace{\sum_{j_g = 1}^{p^g} \boldsymbol{f}^{\{j_g\}}(\boldsymbol{x}) \circ \boldsymbol{z}_{j_g}}_{\text{group-specific effects}} + \underbrace{\sum_{j_g<k_g}^{p_g} \boldsymbol{f}^{\{j_g, k_g\}}(\boldsymbol{x}) \circ (\boldsymbol{z}_{j_g} \circ \boldsymbol{z}_{k_g})}_{\text{second order group effects}}, \quad \text{where}\\
    \boldsymbol{f}^{\emptyset}(\boldsymbol{x}) &= \boldsymbol{f}_{\emptyset}^{\emptyset} + \sum_{j= 1}^p \boldsymbol{f}_{\{j\}}^{\emptyset}(\boldsymbol{x}_j) + \sum_{j<k}^p \boldsymbol{f}_{\{j, k\}}^{\emptyset}(\boldsymbol{x}_j, \boldsymbol{x}_k) + \dots,\\
    \boldsymbol{f}^{j_g}(\boldsymbol{x}) &=  \boldsymbol{f}_{\emptyset}^{j_g} + \sum_{j = 1}^p \boldsymbol{f}_{\{j\}}^{j_g}(\boldsymbol{x}_j) + \sum_{j<k}^p \boldsymbol{f}_{\{j, k\}}^{j_g}(\boldsymbol{x}_j, \boldsymbol{x}_k) + \dots ,\\
    \boldsymbol{f}^{j_g, k_g}(\boldsymbol{x}) &= \boldsymbol{f}_{\emptyset}^{j_g, k_g} + \sum_{j= 1}^p \boldsymbol{f}_{\{j\}}^{j_g, k_g}(\boldsymbol{x}_j) + \sum_{j<k}^p \boldsymbol{f}_{\{j, k\}}^{j_g, k_g}(\boldsymbol{x}_j, \boldsymbol{x}_k) + \dots,
\end{align*}
where $\boldsymbol{f}^{\emptyset}$ represents the global effects across all the groups, with intercept as $\boldsymbol{f}_{\emptyset}^{\emptyset}$, the global main effect of $\boldsymbol{X}_j$ as $\boldsymbol{f}_{\{j\}}^{\emptyset}(\boldsymbol{x}_j)$, and the two-way effects of $\boldsymbol{X}_j, \boldsymbol{X}_k$ as $\boldsymbol{f}_{\{j, k\}}^{\emptyset}(\boldsymbol{x}_j, \boldsymbol{x}_k)$; $\boldsymbol{f}^{j_g}(\boldsymbol{x})$ represents the deviation of group $j_g$ from the global effects, with intercept $\boldsymbol{f}_{\emptyset}^{j_g}$, the main effect of $\boldsymbol{X}_j$ for group $j_g$ as $\boldsymbol{f}_{\{j\}}^{j_g}(\boldsymbol{x}_j)$, and the second order effects of $\boldsymbol{X}_j, \boldsymbol{X}_k$ for group $j_g$ as $\boldsymbol{f}_{\{j, k\}}^{j_g}(\boldsymbol{x}_j, \boldsymbol{x}_k)$; $\boldsymbol{f}^{j_g, k_g}$ represents the additional deviation due to the interaction between groups $i_g$ and $j_g$, similarly with intercept $\boldsymbol{f}_{\emptyset}^{j_g, k_g}$, the one-way effect of $\boldsymbol{X}_j$ as $\boldsymbol{f}_{\{j\}}^{j_g, k_g}(\boldsymbol{x}_j)$ , and the two-way effects of $\boldsymbol{X}_j, \boldsymbol{X}_k$ as $\boldsymbol{f}_{\{j, k\}}^{j_g, k_g}(\boldsymbol{x}_j, \boldsymbol{x}_k)$. To summarize, for each $\boldsymbol{f}^{V_g}$, where $V_g \subseteq [p_g]$ and $|V_g| \le 2$, we decompose it to the summation of identifiable and interpretable lower-order effects through considering the RKHS for $\boldsymbol{f}_{V}^{Vg}$ as
\begin{align*}
    \mathcal{H}_V^{V_g} &= \{\boldsymbol{f}_V^{V_g}: \forall A\subsetneq V, \forall \boldsymbol{f}_A^{V_g}, \langle\boldsymbol{f_V^{V_g}}, \boldsymbol{f_A^{V_g}}\rangle = 0\},
\end{align*}
where $\boldsymbol{f}_V^{V_g}$ are all square-integrable functions of $\boldsymbol{X}_V$ (with respect to its probability measure). Within each level of group-heterogeneity, this orthogonality constraint allows us to focus on the higher-order interaction effects between covariates that has not been explained by all of their lower-order interactions.

For the reproducing kernel $K^{V_g}$ of the $\mathcal{H}^{V_g}$ that $\boldsymbol{f}^{V_g}$ lies in, we similarly consider a sparsity-inducing weighted summation of kernels corresponding to the lower-order effects:
\begin{equation*}
    K^{V_g}(\boldsymbol{x}, \tilde{\boldsymbol{x}}) = \theta_{g, V_g} \sum_{V} \eta^2_{|V|} \prod_{j \in V} \kappa_j K_V(\boldsymbol{x}, \tilde{\boldsymbol{x}}).
\end{equation*}
In order to relax the restrictive assumption of shared importance $\kappa_j$ for $\boldsymbol{X}_j$ across all orders of interactions involved, we introduce more flexibility in modeling interactions that aims to allow for the two following scenarios: (1) the interactions follow weak-hierarchy assumption or no assumption for hierarchy at all; (2) the interactions follow strong-hierarchy assumption, but only a subset of covariates involved in main effects are also involved in higher-order interactions.

Specifically, we introduce a new parameter matrix $\boldsymbol{\tau} \in \mathcal{R}^{p \times Q}$ which characterizes differential variable importance for effects of different orders. This replaces $\kappa_i$ with $\kappa_i \tau_{i, |V|}$ for $\boldsymbol{X}_i$ when examining $|V|$-way interactions involving $\boldsymbol{X}_i$. With this, the kernel becomes
\begin{equation*}
    K^{V_g}(\boldsymbol{x}, \tilde{\boldsymbol{x}}) = \theta_{g, V_g} \sum_{V} \eta^2_{|V|} \prod_{i \in V} \kappa_j \tau_{j, |V|} K_V(\boldsymbol{x}, \tilde{\boldsymbol{x}}),
\end{equation*}
where $\theta_{g, V_g} \in [0, \infty)$ can be similarly interpreted  as the ``scaling factor" that scales the relative contribution of each $\boldsymbol{f}^{V_g}$ to the overall $\boldsymbol{f}$. Larger $\theta_{g, V_g}$ means more deviation of group-specific effects $\boldsymbol{f}^{V_g}$ from the global effects, indicating stronger group heterogeneity; while when $\theta_{g, V_g} = 0,$ for all $V_g \subseteq [p_g],\ V_g \neq \emptyset $, the same global effects apply to all groups, which means no group heterogeneity at all.

To visualize the impact of order-specific variable importance specified through the kernel structure, we plot random draws from a Gaussian process specified by $\sum_{V_g}K^{V_g}$ for a two-dimensional $\boldsymbol{X}$ without group heterogeneity: $\theta_{g, V_g} = 0$ for all $V_g \subseteq [p_g],\ V_g \neq \emptyset$ and $V_{g, \emptyset} = 1$. We consider three different levels $\{1, 0.2, 0\}$ of one-way and two-way effects of $\boldsymbol{X}_2$ specified via $\tau_{2,1}$ and $\tau_{2,2}$ respectively. As shown in Figure \ref{fig:tau_plot}, the structural variation along the $\boldsymbol{X}_2$ direction decreases with $\tau_{2,1}$ with no variation along $\boldsymbol{X}_2$ for $\tau_{2,1}=0$, while the two-way effects remain the same; similarly, decreasing structural variation is observed for two-way effects as $\tau_{2,2}$ decreases, while the one-way effect remains the same.

\begin{figure}[!htbp]
    \centering
    \includegraphics[width=0.85\textwidth]{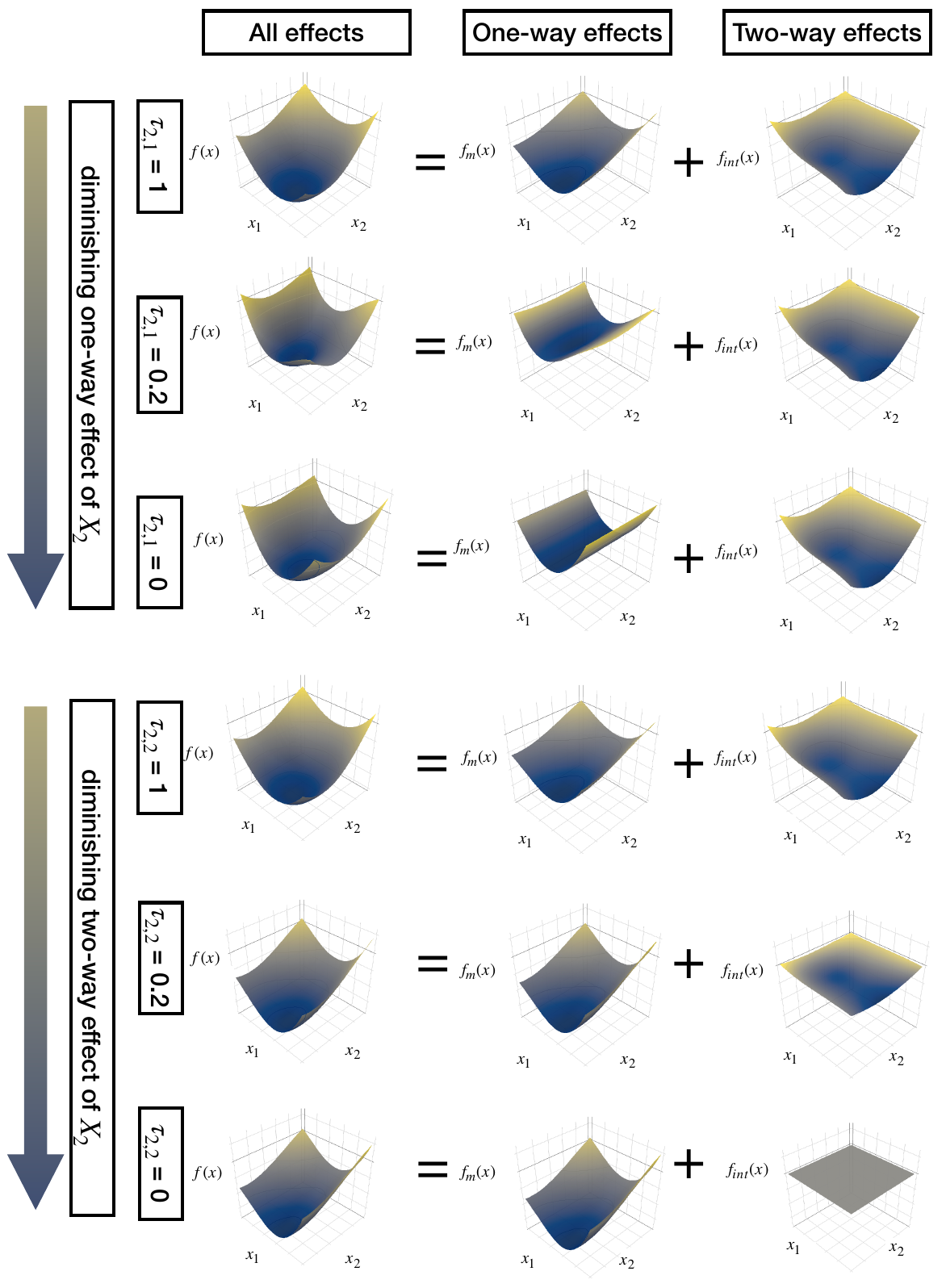}
    \caption{Illustration of the impact of diminishing first-order variable importance $\tau_{2,1}$ and second-order variable importance $\tau_{2,2}$ for two-dimensional $\boldsymbol{X}$ on all, one-way (main) and two-way (second order) effects. For the top three rows, decreasing first-order variable importance $\tau_{2,1} = \{1, 0.2, 0\}$ is applied for $\boldsymbol{X}_2$. Decreasing structural variation is observed along the $\boldsymbol{X}_2$ direction, with no variation at all for $\tau_{2,1} = 0$, for all and one-way effects, while two-way effects remain the same; For the bottom three rows, decreasing second-order variable importance $\tau_{2,2} = \{1, 0.2, 0\}$ is applied for $\boldsymbol{X}_2$. Decreasing structural variation is observed, with no variation for $\tau_{2,2} = 0$, for all and two-way effects, while one-way effects remain the same.}
    \label{fig:tau_plot}
\end{figure}

Similarly, to avoid combinatorial summations over the number of groups and allow heterogeneity induced by a group to vary across different orders of interactions between the group design matrix $\boldsymbol{Z}$, $\theta_{g, V_g}$ can be reparametrized as $\theta_{g, V_g} = \eta^2_{g, |V_g|} \prod_{j_g \in V_g} \kappa_{g, j_g}\cdot \tau_{g, j_g,  |V_g|}$. Thus, our proposed kernel is
\begin{align}
    & K_{\text{HARNESS}}((\boldsymbol{x, z}), (\tilde{\boldsymbol{x}}, \tilde{\boldsymbol{z}})) \nonumber \\ &= \underbrace{\sum_{V_g} \eta^2_{g, |V_g|} \prod_{j_g \in V_g} \kappa_{g, j_g}\cdot \tau_{g, j_g, |V_g|}\cdot z_{j_g} \tilde{z}_{j_g}}_{\text{\normalsize{\textcircled{2}: group heterogeneity components}}} \underbrace{\left\{ \sum_{V} \eta^2_{|V|} \prod_{j \in V} \kappa_j \tau_{j, |V|} K_V(\boldsymbol{x}, \tilde{\boldsymbol{x}}) \right\}}_{\text{\normalsize{\textcircled{1}: base kernel}}} \label{eq:kernel_order},\\
    &= \underbrace{\eta^2_{g, 0} \left\{ \sum_{V} \eta^2_{|V|} \prod_{j \in V} \kappa_j \tau_{j, |V|} K_V(\boldsymbol{x}, \tilde{\boldsymbol{x}})\right\}}_{\text{\normalsize{common kernel structure}}}\notag\\
    &\quad + \underbrace{\eta^2_{g, 1}\sum_{j_g} \kappa_{g, j_g}\cdot \tau_{g, j_g, 1} \cdot z_{j_g} \tilde{z}_{j_g}\left\{ \sum_{V} \eta^2_{|V|} \prod_{j \in V} \kappa_j \tau_{j, |V|} K_V(\boldsymbol{x}, \tilde{\boldsymbol{x}})\right\}}_{\text{\normalsize{heterogeneity induced by main effects of groups}}}\notag\\
    &\quad + \underbrace{\eta^2_{g, 2} \sum_{|V_g| = 2} \prod_{j_g\in V_g} \kappa_{g, j_g}\cdot  \tau_{g, j_g, 2}\cdot z_{j_g} \tilde{z}_{j_g} \left\{ \sum_{V} \eta^2_{|V|} \prod_{j \in V} \kappa_j \tau_{j, |V|} K_V(\boldsymbol{x}, \tilde{\boldsymbol{x}})\right\}}_{\text{\normalsize{heterogeneity induced by second order effects of groups}}}\notag,
\end{align}
where $\kappa_{g, i_g} \in [0, \infty)$ 
measures the overall heterogeneity of group $i_g$, which is composed of heterogeneity induced by its $|V_g|$-way effects measured by $\tau_{g, i_g, |V_g|} \in [0, \infty)$}. The parameter $\eta^2_{g, |V_g|} \in [0, \infty)$ modulates the overall strength of heterogeneity due to $|V_g|$-way interactions. 
Specifically, with two-level group structure, $\eta^2_{g, 0}$ measures the importance of the common kernel structure across all groups in function estimation; $\eta^2_{g, 1}$ accounts for the overall heterogeneity of main effects of groups; $\eta^2_{g, 2}$ accounts for the overall importance of the heterogeneity induced by the two-way effects of groups, which simply includes interactions between each level-1 group and all level-2 groups nested within it. 

To aid in the interpretation of $\boldsymbol{f}$, we describe the RKHS $\mathcal{H}$ corresponding to the $K_{\text{HARNESS}}$ as a hierarchical direct sum that closely follows the kernel structure: within each group, the space is decomposed as a direct sum over covariates, and the overall RKHS is then formed as a direct sum across all groups as
\begin{equation*}
    \mathcal{H} = \oplus_{|V_g|\le Q^g}(\mathcal{H}'^{V_g}) = \oplus_{|V_g|\le Q^g}\left\{\oplus_{V} (\mathcal{H}_V'^{V_g})\right\},
\end{equation*}
where $\mathcal{H}'^{V_g}$ is direct sum across $V$ of the corresponding RKHS $\mathcal{H}_V'^{V_g}$ for the reproducing kernel $\eta^2_{|V|} \prod_{j \in V} \kappa_j \tau_{j, |V|}  K_V$; and $\mathcal{H}$ is the direct sum across $V_g$ of $\mathcal{H}'^{V_g}$, which corresponds to the reproducing kernel of $\mathcal{H}^{V_g}$ scaled by the group-heterogeneity importance measure $\eta^2_{g, |V_g|} \prod_{j_g \in V_g} \kappa_{g, j_g}\cdot \tau_{g, j_g,  |V_g|}$.

To summarize, $K_{\text{HARNESS}}$ builds upon the principle of sparsity-inducing kernel summation for variable selection, while further incorporating an order-specific measure of variable importance and a group-specific kernel structure. These enhancements enable the model to account for potential inconsistencies in effect orders and heterogeneity across groups.

To visualize the impact of group heterogeneity specified through the kernel structure, we similarly plot random draws from a Gaussian process specified by $K_{\text{HARNESS}}$ with three levels of group heterogeneity. Here for simplicity we only consider having two level-1 groups, the $\eta_1$ for small and large group effect is chosen correspondingly as 0.25 and 1. 

\begin{figure}[!htbp]
    \centering
    \includegraphics[width=1\textwidth]{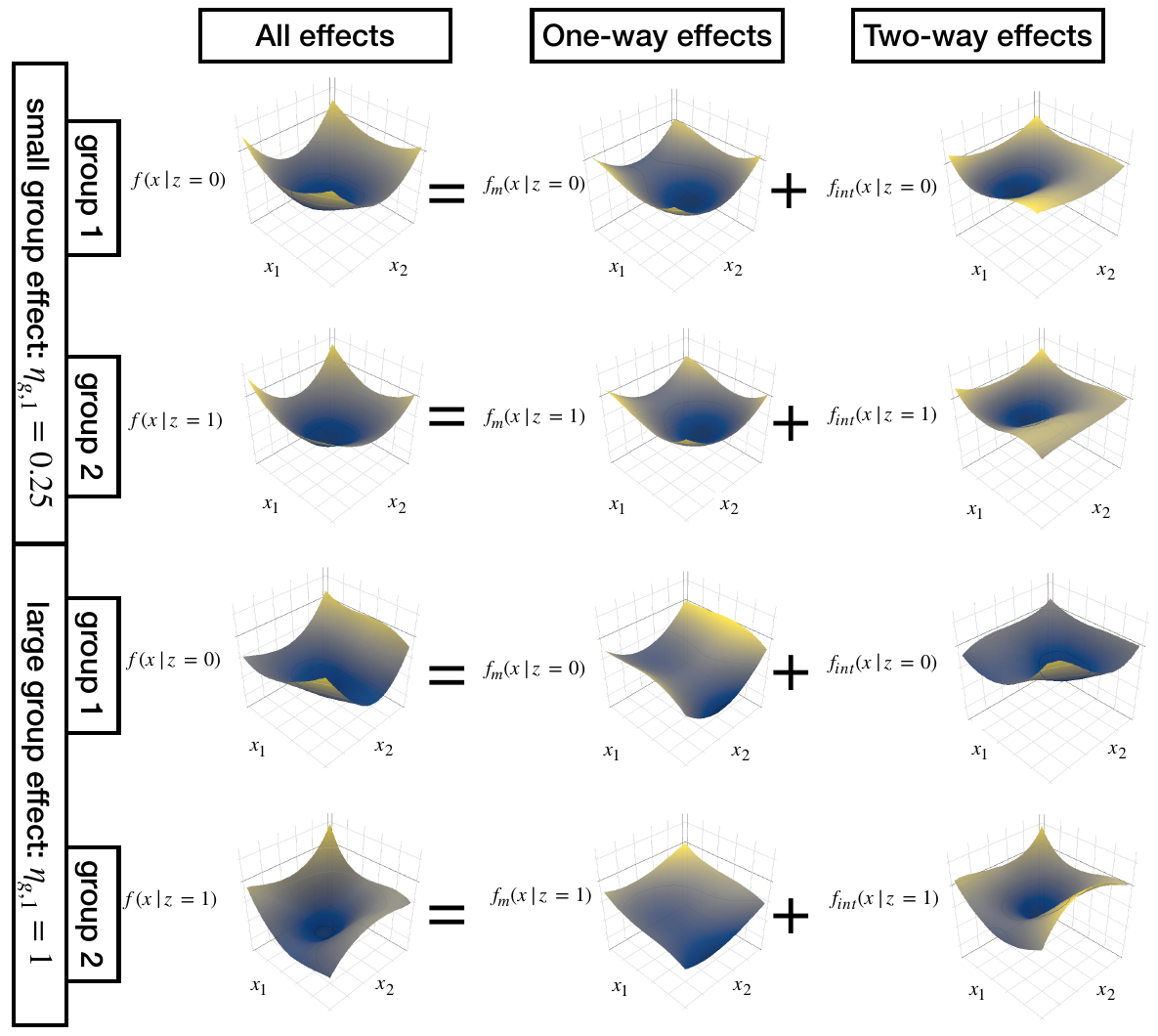}
    \caption{Illustration of the impact of increasing group heterogeneity for two-dimensional $\boldsymbol{X}$ with two level-1 groups on all, one-way and two-way effects. When small group effect $\eta_{g,1} = 0.25$ is applied (top two rows), minimal structural inconsistency between group 1 and group 2 is observed for all, one-way (main) and two-way (second order) effects; when large group effect $\eta_{g,1} = 1$ is applied (bottom two rows), more substantial structural inconsistency is observed for all, one-way and two-way effects.}
    \label{fig:3d_group}
\end{figure}

As demonstrated in Figure~\ref{fig:3d_group}, increasing structural discrepancy is observed for all effects between the two groups as the group effect increases. Additionally, the same trend observed for one-way and two-way effects demonstrates that the group heterogeneity can be decomposed to have impacts on one-way and two-way effects separately.

\section{Implementation and Final Algorithm}

\subsection{Kernel Evaluation}\label{sec:implement}

$K_{\text{HARNESS}}$ involves recursive summation over $p^Q + p_g ^{Q_g}$ distinct interaction pairs, with $Q$ defined as the highest order of interest within covariates $\boldsymbol{X}$. This is computationally intensive in practice given the high-dimensionality of EHR data. Under the covariate independence assumption that $\mu = \mu_{\otimes},$ where $\boldsymbol{X}_i \overset{\text{i.i.d.}}{\sim} \mu$ and $\mu_{\otimes}(\boldsymbol{x}) := \bigotimes_{j=1}^p \mu_j(\boldsymbol{x}_j)$ with $\mu_j$ as the marginal distribution of $\boldsymbol{X}_j$, we have $k_V = \prod_i k_i$ \citep{gu_smoothing_1993}, upon which Theorem 2 of \citet{agrawal_skim-fa_2023} can be used for efficient evaluation of part \textcircled{1} of Equation \eqref{eq:kernel_order} through the following trick:
\begin{align*}
    &\text{\textcircled{1}} = \sum_{q=1}^Q \eta^2_q \overline{k}_q(\boldsymbol{x}, \boldsymbol{\Tilde{x}}) \quad \text{s.t.}\\
    &\overline{k}_q(\boldsymbol{x}, \boldsymbol{\Tilde{x}}) = \frac{1}{q}\sum_{s=1}^q (-1)^{s+1}\overline{k}_{q-s}(\boldsymbol{x}, \boldsymbol{\Tilde{x}}), \quad \overline{k}_0(\boldsymbol{x}, \boldsymbol{\Tilde{x}}) = 1, k^s(\boldsymbol{x}, \boldsymbol{\Tilde{x}}) = \sum_{i=1}^p (\kappa_i\tau_{i, |V|})^{2s}[k_i(\boldsymbol{x}, \boldsymbol{\Tilde{x}})]^2.
\end{align*}
Thus, for the case when $Q=2$, we have part \textcircled{1} of $K(\boldsymbol{x}, \boldsymbol{\Tilde{x}})$ as
\begin{equation*}
    0.5\eta_2^2\left\{(\sum_{i = 1}^p (\kappa_i\tau_{i, 2})^2 k_i(x_i, \Tilde{x_i})^2 - \sum_{i=1}^p (\kappa_i\tau_{i, 2})^4[k_i(x_i, \Tilde{x_i})]^2\right\} + \eta_1^2 \sum_{i = 1}^p (\kappa_i\tau_{i, 1}) ^2 k_i (x_i, \Tilde{x_i}) + \eta_0 ^2,
\end{equation*}
and utilizing the same trick, the part \textcircled{2} can be evaluated as 
\begin{equation*}
    0.5\eta_{g,2}^2\left\{(\sum_{i_g = 1}^{p_g} (\kappa_{g, i_g}\tau_{g, i_g, 2})^2 z_i \Tilde{z_i} - \sum_{i_g=1}^{p_g} (\kappa_{g, i_g}\tau_{g, i_g, 2})^4z_i \Tilde{z_i}\right\} + \eta_{g,1}^2 \sum_{i_g = 1}^{p_g} (\kappa_{g, i_g}\tau_{g, i_g, 1}) ^2 z_i \Tilde{z_i} + \eta_{g, 0} ^2.
\end{equation*}
Intuitively, the first component for both part \textcircled{1} and part \textcircled{2} representing the second-order interactions can be interpreted as a combinatorial expression involving the difference between the square of a sum and the sum of squares, keeping pairwise interactions between distinct covariates only. 
We adopt this kernel formulation despite the potential violation of covariate independence assumption for real data, as there exists a unique change-of-basis which maps this representation to the general case $\mu \neq \mu_{\otimes}$ (Theorem 3 of \citet{agrawal_skim-fa_2023}), allowing us to retain performance while simplifying the framework.

\subsection{Function Estimation}
We consider the use of cross-validation to choose $\hat{\boldsymbol{\xi}}$ and $\hat{\boldsymbol{\xi}_g}$ of hyperparameters $\boldsymbol{\xi} = (\boldsymbol{\kappa}\in \mathcal{R}^p, \boldsymbol{\tau}\in \mathcal{R}^{p \times (Q+1)}, \boldsymbol{\eta} \in \mathcal{R}^{Q+1})$ and $\boldsymbol{\xi}_g = (\boldsymbol{\kappa}_{g}\in \mathcal{R}^{p_g}, \boldsymbol{\tau}_g\in \mathcal{R}^{p_g \times (Q_g + 1)}, \boldsymbol{\eta}_g \in \mathcal{R}^{Q_g+1})$ characterizing the kernel. Specifically using the general approach of \citet{agrawal_skim-fa_2023}, we choose hyperparameters by minimizing the leave-$m$-out cross validation loss as
\begin{align*}
    L(\boldsymbol{\xi}, \boldsymbol{\xi}_g) &= \frac{1}{\binom{n}{m}} \sum_{\substack{A: A \subset [n] \\ |A| = n - m}} \left[ - \sum_{i \in [n] \setminus A}\log\left\{\sigma \left(Y^{(i)}\boldsymbol{f}_A(\boldsymbol{x}^{(i)}, \boldsymbol{z}^{(i)})\right)\right\}  \right],
\end{align*}
where $\boldsymbol{f}_A(\boldsymbol{x}^{(i)}, \boldsymbol{z}^{(i)})$ is the kernel logistic regression fit of Equation \eqref{eq:obj} using the subset of data in $A$. We treat data in $A$ as the training data through which the function is estimated, and then evaluate and minimize the loss on the testing data in $[n] \setminus A$.

We consider to choose hyperparameters through stochastic gradient-descent for scalability to large data, with details provided in Section \ref{sec:large}. For practical purpose, we do not directly use the cross-validation loss $L(\boldsymbol{\xi}, \boldsymbol{\xi}_g)$ but instead approximate it by the loss for a single draw $A^{(t)} \sim \pi$, where $\pi$ is the uniform distributions over all $n-m$ subsets of $[n]$, as
\begin{equation}\label{eq:loss_iter}
    L^{(t)}(\boldsymbol{\xi}, \boldsymbol{\xi}_g) = - \sum_{i \in [n] \setminus A^{(t)}} \log\left\{\sigma\left(Y^{(i)}\boldsymbol{f}_{A^{(t)}}(\boldsymbol{x}^{(i)}, \boldsymbol{z}^{(i)})\right)\right\}.
\end{equation}
To estimate $\boldsymbol{f}_{A^{(t)}}$, we consider maximizing $\Psi(\boldsymbol{f}_{A^{(t)}})$ defined as the negative of the objective function specified in Equation \eqref{eq:obj}. Specifically,
\begin{align*}
    \Psi(\boldsymbol{f}_{A^{(t)}}) &\equiv \log p(Y_{A^{(t)}}|\boldsymbol{f}_{A^{(t)}}) - \lambda ||\boldsymbol{f}_{A^{(t)}}||^2_{\mathcal{H}},\quad\text{where}\\
    ||\boldsymbol{f}_{A^{(t)}}||^2_{\mathcal{H}} &= \alpha^TK_{A^{(t)}}\alpha = (K_{A^{(t)}}^{-1}\boldsymbol{f}_{A^{(t)}})^TK_{A^{(t)}}(K_{A^{(t)}}^{-1}\boldsymbol{f}_{A^{(t)}}) = \boldsymbol{f}_{A^{(t)}}^TK_{A^{(t)}}^{-1}\boldsymbol{f}_{A^{(t)}},
\end{align*}
and $Y_{A^{(t)}} = \{Y_i\}_{i \in A^{(t)}}$ is the outcome for training set, and $K_{A^{(t)}}$ denotes the HARNESS kernel constructed using the training data $(\boldsymbol{X}_{A^{(t)}}, \boldsymbol{Z}_{A^{(t)}}) = (\{\boldsymbol{X}_i\}_{i \in A^{(t)}}, \{\boldsymbol{Z}_i\}_{i \in A^{(t)}})$ as $K_{A^{(t)}} = K_{\text{HARNESS}}((\boldsymbol{x}_{A^{(t)}}, \boldsymbol{z}_{A^{(t)}}), (\boldsymbol{\tilde{x}}_{A^{(t)}}, \boldsymbol{\tilde{z}}_{A^{(t)}}))$ for simplicity.

We use Newton's method as described in \citet{rasmussen_gaussian_2008} to find the maximum of $\Psi$. Specifically with the iteration
\begin{align*}
    \boldsymbol{f}_{A^{(t)}}^{\text{new}} &= \boldsymbol{f}_{A^{(t)}} - \left\{\nabla \nabla \Psi(\boldsymbol{f}_{A^{(t)}})\right\}^{-1} \nabla\Psi(\boldsymbol{f}_{A^{(t)}})  \quad\text{where},\\
    \nabla\Psi(\boldsymbol{f}_{A^{(t)}}) &= \nabla \log p(Y_{A^{(t)}}|\boldsymbol{f}_{A^{(t)}}) - 2\lambda K_{A^{(t)}}^{-1}\boldsymbol{f}_{A^{(t)}},\\
    \nabla \nabla \Psi(\boldsymbol{f}_{A^{(t)}}) &= \nabla \nabla \log p(Y_{A^{(t)}}|\boldsymbol{f}_{A^{(t)}}) - 2\lambda K_{A^{(t)}}^{-1} = -W_{A^{(t)}} - 2\lambda K_{A^{(t)}}^{-1}, 
\end{align*}
and $W_{A^{(t)}} \equiv - \nabla \nabla \log p(Y_{A^{(t)}}|\boldsymbol{f}_{A^{(t)}})$ is diagonal. In particular, when the sigmoid function is logistic, we have $\frac{\partial}{\partial f_i}\log p(Y_i | f_i) = (Y_i + 1)/ 2 - p_i$; and $[W_{A^{(t)}}]_{ii} = -p_i(1 - p_i)$ where $p_i = p(y_i = 1 | f_i)$. After finding $\hat{\boldsymbol{f}}_{A^{(t)}}$, we can get the predicted $\hat{\boldsymbol{f}}_{A^{(t)}}(\boldsymbol{x}_{A^{(t)}}^*, \boldsymbol{z}_{A^{(t)}}^*)$ for testing data $(\boldsymbol{X}_{A^{(t)}}^*,  \boldsymbol{Z}_{A^{(t)}}^*) = (\{\boldsymbol{X}_i\}_{i \in [n] \setminus A^{(t)}}, \{\boldsymbol{Z}_i\}_{i \in [n] \setminus A^{(t)}})$ as
\begin{equation}\label{eq:predict}
   \hat{\boldsymbol{f}}_{A^{(t)}}(\boldsymbol{x}_{A^{(t)}}^*, \boldsymbol{z}_{A^{(t)}}^*) = k_{A^{(t)}}^{*T} K_{A^{(t)}}^{-1} \hat{\boldsymbol{f}}_{A^{(t)}}(\boldsymbol{x}_{A^{(t)}}, \boldsymbol{z}_{A^{(t)}}),
\end{equation}
where $k_{A^{(t)}}^* = K_{\text{HARNESS}}((\boldsymbol{x}_{A^{(t)}}^*, \boldsymbol{z}_{A^{(t)}}^*), (\boldsymbol{x}_{A^{(t)}}, \boldsymbol{z}_{A^{(t)}}))$ for simplicity. The predicted $\hat{\boldsymbol{f}}_{A^{(t)}}$ is then used to calculate cross-validation loss, upon which the chosen hyperparameters are updated via stochastic gradient-descent.
\subsection{Scalability to Large Data}\label{sec:large}

Handling large-scale kernels like $K_{A^{(t)}}$ poses a significant computational challenge in predictive modeling, particularly in healthcare applications where large-scale EHR are common. Commonly used methods include global approximation through Nystrom Approximation \citep{hensman_gaussian_2013}, sparse and low-rank approximation \citep{gneiting_compactly_2002} or smart-sample selection \citep{chalupka_framework_2013}; and local approximation through combining models locally optimized for smaller subsets \citep{gramacy_lagp_2016, masoudnia_mixture_2014}. However, forcing low-rank approximations introduces information loss which potentially results in suboptimal performance \citep{liu_when_2020}. Additionally, in practice, we find that local approximation approaches tend to overfit locally on smaller subregions of the data that decreases overall performance. We propose instead to optimize kernel parameters of the model globally through stochastic gradient descent for training step and use D\&R for the final prediction step due to implementation simplicity and superior practical performance.

\subsubsection{Optimization of parameters through stochastic gradient descent}

Consider a randomly selected mini-batch $B_{A^{(t)}} \subset A^{(t)}$ of size $b$ where $b \ll n-m$. Then the loss function defined in Equation \eqref{eq:loss_iter} is approximated as,
\begin{equation}\label{eq:loss_iter_mini}
    L_b^{(t)}(\boldsymbol{\xi}, \boldsymbol{\xi}_g) = - \sum_{i \in [n] \setminus A^{(t)}} \log\left\{\sigma \left(Y^{(i)}\boldsymbol{f}_{B_{A^{(t)}}}(\boldsymbol{x}^{(i)}, \boldsymbol{z}^{(i)})\right)\right\},
\end{equation}

where $\boldsymbol{f}_{B_{A^{(t)}}}$ is estimated similarly through maximizing $\Psi(\boldsymbol{f}_{B_{A^{(t)}}})$ via Newton's method. The only difference is that now only $(\boldsymbol{X}_{B_{A^{(t)}}}, \boldsymbol{Z}_{B_{A^{(t)}}}) = (\{\boldsymbol{X}_i\}_{i \in B_{A^{(t)}}}, \{\boldsymbol{Z}_i\}_{i \in B_{A^{(t)}}})$ is used as the training data. This allows us to construct and invert a kernel matrix of size $b$ instead of $n-m$ for each iteration, which significantly improves the time and memory efficiency for the training process.

\subsubsection{Prediction of out-of-sample data through Divide and Recombine}

After the training model converges, we use D\&R for the prediction on out-of-sample data $\boldsymbol{X}^*, \boldsymbol{Z}^*$ of size $n^*$. Specifically, we evenly and uniformly divide the data $\{(\boldsymbol{X}_1, Y_1), \dots, (\boldsymbol{X}_n, Y_n)\}$ into $D$ disjoint subsets $S_1, \dots, S_D$. To obtain predictions on new data $\boldsymbol{X}^*$, for each $d, d\in 1, \dots, D$, estimate the local $\hat{\boldsymbol{f}}_d(\boldsymbol{x}_d, \boldsymbol{z}_d)$, and then average the local predicted $\boldsymbol{f}^*_{d}(\boldsymbol{x}^*, \boldsymbol{z}^*)$ as
\begin{equation*}
    \hat{\boldsymbol{f}}^*(\boldsymbol{x}^*, \boldsymbol{z}^*) = \frac{1}{D}\sum_{d=1}^D \hat{\boldsymbol{f}}^*_{d}(\boldsymbol{x}^*, \boldsymbol{z}^*) = \frac{1}{D}\sum_{d=1}^D \ k_d^{*T} \boldsymbol{K}_d^{-1} \hat{\boldsymbol{f}_d}(\boldsymbol{x}_d, \boldsymbol{z}_d),
\end{equation*}
where $k_d^* = K_{\text{HARNESS}}((\boldsymbol{x}_d, \boldsymbol{z}_d), (\boldsymbol{x}^*, \boldsymbol{z}^*))$ and $\boldsymbol{K}_d = \boldsymbol{K}_{\text{HARNESS}}((\boldsymbol{x}_d, \boldsymbol{z}_d), (\tilde{\boldsymbol{x}}_d, \tilde{\boldsymbol{z}}_d))$.

Instead of using only a subset of size $b$ from the training data during hyperparmeter optimization step, for the prediction on out-of-sample data we want to make use of the entire data. This is mainly because (1) the hyperparameter optimization step requires the function estimation calculation of the predictive mean for each iteration. This increases the total training time to be $\frac{n}{m}$ times longer, which is prohibitively long for very large datasets; (2) on the other hand, the final prediction step is essentially just ``one iteration" and calculating $D$ predictive mean only increases the time for this ``one iteration" to be $D$ times longer. This extra computation time allows the final predictive mean to be more reliable by averaging across different subsets. Algorithm 1 summarizes the whole hyperparameter optimization and prediction procedure of HARNESS. 

\section{Simulation Studies}
\subsection{Setting and Rationale}

Motivated by the sample size and group structure of the hematologic data, we consider simulating $\boldsymbol{X} \in \mathbb{R}^{25,000 \times 100}$ where each variable $\boldsymbol{X}_j \overset{\text{i.i.d.}}{\sim} \text{Uniform}(-1, 1)$ for $j=1,\dots, 100$. We further consider that $\boldsymbol{X}$ are composed of 5 subgroups with each group having a size of 5,000, with subgroup-specific heterogeneity incorporated into the data-generating model to reflect the heterogeneity observed in the real data. Following the hematologic data structure, we consider that $\boldsymbol{X}$ is made up of 10 ``years'' of data with each year having a size of 2,500, for which year-specific variation is considered in the data-generating model to capture the temporal changes in readmission rates observed in our motivating data. Additionally, we also conduct simulations with smaller sample sizes to evaluate the method's performance in scenarios where it has a higher risk of overfitting, thereby assessing its robustness and generalizability. Specifically, we consider $n=5,000$ with 5 subgroups each of size 1,000. Similarly, we consider it composed of 10-year data with each year size of 500.

\begin{algorithm}[H]
\caption{Heterogeneity-Aware Regression with Nonparametric Estimation and Structured Selection (HARNESS) Algorithm}\label{alg:gradient_descent}
\begin{algorithmic}[1]
    \State \textbf{Inputs:} the main design matrix $\boldsymbol{X}_{n \times p}$, the group design matrix $\boldsymbol{Z}_{n \times p_g}$, the outcome $Y_{n \times 1}$, the main design matrix $\boldsymbol{X}^*_{n^* \times p}$ and the group design matrix $\boldsymbol{Z}^*_{n^* \times p_g}$ for the out-of-sample data; $b$ for mini-batch size; $D$ for the number of disjoint subsets for D\&R.
    
    \State \textbf{Initialize:}\\
        $\boldsymbol{\xi}^{(0)} = (\boldsymbol{\kappa}^{(0)}\in \mathcal{R}^p, \boldsymbol{\tau}^{(0)}\in \mathcal{R}^{p \times Q}, \boldsymbol{\eta}^{(0)} \in \mathcal{R}^{Q+1})$ \Comment{Learnable parameters for main design matrix}\\
        $\boldsymbol{\xi}_g^{(0)} = (\boldsymbol{\kappa}_{g}^{(0)}\in \mathcal{R}^{p_g}, \boldsymbol{\tau}_g^{(0)}\in \mathcal{R}^{p_g \times Q_g}, \boldsymbol{\eta}_g^{(0)} \in \mathcal{R}^{Q_g+1})$ \Comment{Learnable parameters for group design matrix}
            \For{$t = 0$ to $T$}
                \State Randomly select $A^{(t)} \sim \pi$, where $\pi$ is the uniform distributions over all $n-m$ subsets of $[n]$. Randomly select $B_{A^{(t)}} \subset A^{(t)}$ of size $b$ as the mini-batch set: $\boldsymbol{X}_{B_{A^{(t)}}}, \boldsymbol{Z}_{B_{A^{(t)}}}$, and $Y_{B_{A^{(t)}}}$; the remaining $n-m$ samples $\boldsymbol{X}_{A^{(t)}}^*, \boldsymbol{Z}_{A^{(t)}}^*$, and $Y_{A^{(t)}}^*$ is the inner testing set 
                
                \State Compute the HARNESS kernel $K_{A^{(t)}} = K_{\text{HARNESS}}((\boldsymbol{x}_{B_{A^{(t)}}}, \boldsymbol{z}_{B_{A^{(t)}}}), (\boldsymbol{\tilde{x}}_{B_{A^{(t)}}}, \boldsymbol{\tilde{z}}_{B_{A^{(t)}}}))$ as defined in Eq. \ref{eq:kernel_order} 
                \State Compute $\hat{\boldsymbol{f}}_{B_{A^{(t)}}}(\boldsymbol{x}_{B_{A^{(t)}}}, \boldsymbol{z}_{B_{A^{(t)}}})$ via Newton's method
                \State Compute the predicted $\hat{\boldsymbol{f}}^*_{B_{A^{(t)}}}(\boldsymbol{x}_{A^{(t)}}^*, \boldsymbol{z}_{A^{(t)}}^*)$ for the inner testing set via Eq. \ref{eq:predict}
                \State Get the loss $L_b^{(t)}(\boldsymbol{\xi}^{(t)}, \boldsymbol{\xi}_g^{(t)})$ as defined in Eq. \ref{eq:loss_iter_mini}
                \State Take stochastic gradient descent step $(\boldsymbol{\xi}^{(t+1)}, \boldsymbol{\xi}_g^{(t+1)}) = (\boldsymbol{\xi}^{(t)}, \boldsymbol{\xi}_g^{(t)}) - \gamma \nabla_{\boldsymbol{\xi}^{(t)}, \boldsymbol{\xi}_g^{(t)}} L_b^{(t)}$ \Comment{$\gamma$ is the hyperparameter for learning rate; the sparsity-inducing priors for $\boldsymbol{\kappa}$ and $\boldsymbol{\tau}$ are chosen as the same as \citet{agrawal_skim-fa_2023} to take stochastic gradient descent.}
            \EndFor
    \State \textbf{Outputs:} $\boldsymbol{\xi}^{(T)}, \boldsymbol{\xi}_g^{(T)}$ \Comment{Optimized parameters for main design matrix and group design matrix respectively.}
    
    \State \textbf{Prediction:}
        \State Uniformally and evenly divide the data $\{(\boldsymbol{X}_1, Y_1), \dots, (\boldsymbol{X}_n, Y_n)\}$ into $D$ disjoints subsets $S_1, \dots, S_D$.
            \For{$d = 1$ to $D$}
                \State Estimate $\hat{\boldsymbol{f}_d}$ similarly through first computing the HARNESS kernel with parameters $\boldsymbol{\xi}^{(T)}, \boldsymbol{\xi}_g^{(T)}$  and then using Newton's method
                \State Get the predicted $\hat{\boldsymbol{f}}^*_{d}(\boldsymbol{x}^*, \boldsymbol{z}^*)$ for testing set
            \EndFor
            \State $\hat{\boldsymbol{f}}^* = \frac{1}{D}\sum_{d=1}^D \hat{\boldsymbol{f}}^*_{d}$
    \State \textbf{Output:} $\hat{\boldsymbol{f}}^*$
    
\end{algorithmic}
\end{algorithm}

For both sample sizes, we assume that only the first 20 variables of $\boldsymbol{X}$ are involved in the one-way effects: $\boldsymbol{X}_{1:6}$ has a linear relationship with $Y$; $\boldsymbol{X}_{7:13}$ has a sinusoidal relationship with $Y$; and $\boldsymbol{X}_{14:20}$ has a quadratic relationship with $Y$. 15 pairs of interactions are also added as specified in the set $S$, with the choice of interaction pairs $S$ different for each simulation settings, explained in detail later. 

To incorporate group heterogeneity and differences across years, we consider to generate ``base coefficients" for each relationship and disperse them by groups and then by years. Specifically, the base coefficients are $\alpha_j \overset{\text{i.i.d.}}{\sim} \mathcal{N}(1, 0.1), j = 1, \dots, 6$ for the linear relationship; $\beta_k \overset{\text{i.i.d.}}{\sim} \mathcal{N}(5, 0.1), k = 7, \dots, 13$ for the sinusoidal relationship; $\gamma_l \overset{\text{i.i.d.}}{\sim} \mathcal{N}(1, 0.1), l = 14, \dots, 20$ for the quadratic relationship; and $\zeta_r \overset{\text{i.i.d.}}{\sim} \mathcal{N}(1, 0.1), r = 1, \dots, 15$ for the interaction pairs. Then for any base coefficient $b \in \{\boldsymbol{\alpha}, \boldsymbol{\beta}, \boldsymbol{\gamma}, \boldsymbol{\zeta}\}$, we disperse it first by group such that for each group $g$, $b^g \overset{\text{i.i.d.}}{\sim} \mathcal{N}(b, 5)$; and further for each year $t$ within group $g$, we further disperse it as $b^{g, t} \overset{\text{i.i.d.}}{\sim} \mathcal{N}(b^g, 0.1)$. To summarize, the outcome $y$ is generated as $Y \sim \text{Bernoulli}(\pi(\boldsymbol{f}))$, with
\begin{align}\label{eq:simulation}
    \boldsymbol{f} = \sum_{t=1}^{10} \sum_{g=1}^5 \Bigg\{\underbrace{\sum_{j=1}^6 \alpha_j^{g, t} \boldsymbol{x}_j^{g, t}}_{\mathclap{\substack{\text{\textcircled{1}linear one-way} \\ \text{effects}}}} +  \underbrace{\sum_{k=7}^{13} \sin{(\beta_k^{g, t} \boldsymbol{x}_k^{g, t})}
    + \sum_{l = 14}^{20} \gamma_l^{g, t} (\boldsymbol{x}_l^{g, t})^2}_{\text{\textcircled{2}nonlinear one-way effects}} + \underbrace{\sum_{r = 1}^{15} \zeta_r^{g,t} \boldsymbol{x}_{S_r[1]} \boldsymbol{x}_{S_r[2]}}_{\text{\textcircled{3}two-way effects}}\Bigg\},
\end{align}
where $(S_r[1], S_r[2])$ represents the $r$-th pair from the set $S$ for all interaction pairs we consider to include in a particular simulation setting. 
Dispersing coefficients in this manner emphasizes group-level heterogeneity while allowing for smaller temporal fluctuations within each group. This setup prioritizes cross-group variation, incorporates temporal shifts without assuming any specific trend over time, and enables a more realistic assessment of the method’s prospective performance on unforeseen future data.

The simulation settings are motivated by the patterns observed from the hematologic data and key assumptions of the methodology. The comparison methods considered and justifications are illustrated in Table \ref{tab:method}. HARNESS (w/o group or order-specificity), without the introduction of order-specific $\kappa$, encourages agreement between the covariates involved in main effects and higher-order effects. To explore the effectiveness of order-specific variable importance introduced in HARNESS, we consider three different levels of agreement between main effects and higher-order effects, all of which are under strong-hierarchy assumption as demonstrated in hematologic results. Additionally, while the kernel is able to handle non-linear and higher-order effects, theoretically it is also able to adapt to the case where only one-way linear relationships are present. We therefore consider four simulation settings by adjusting each part of Equation \eqref{eq:simulation} as summarized in Table \ref{tab:sim_setting}.

\begin{table}[]
\caption{Methods considered for performance evaluation and rationale for inclusion.}
\label{tab:method}
\resizebox{\columnwidth}{!}{%
\begin{tabular}{@{}lll@{}}
\toprule
Name &
  Description &
  Rationale for Inclusion \\ \midrule
HARNESS &
  \begin{tabular}[c]{@{}l@{}}Heterogeneity-Aware Regression with \\ Nonparametric Estimation and Structured Selection.\end{tabular} &
  Proposed method. \\ \midrule
\begin{tabular}[c]{@{}l@{}}HARNESS \\ (w/o group or order-specificity)\end{tabular} &
  \begin{tabular}[c]{@{}l@{}}A version of HARNESS that does not include the \\ order-specific variable importance and group \\ heterogeneity.\end{tabular} &
  Evaluate the effectiveness of newly introduced methodology. \\ \midrule
\begin{tabular}[c]{@{}l@{}}HARNESS \\ (w/o order-specificity)\end{tabular} &
  \begin{tabular}[c]{@{}l@{}}A version of HARNESS that includes only the \\ group heterogeneity.\end{tabular} &
  \begin{tabular}[c]{@{}l@{}} Evaluate the effectiveness of introducing\\ order-specific variable importance.\end{tabular} \\ \midrule
Lasso group &
  Lasso regression implemented on $\boldsymbol{X} \otimes \boldsymbol{Z}$. &
  \begin{tabular}[c]{@{}l@{}}Broad application in readmission prediction modeling; \\ high interpretability.\end{tabular} \\ \midrule
XGBoost &
  \begin{tabular}[c]{@{}l@{}}Extreme Gradient Boosting with hyperparameters \\ determined through 5-fold cross-validation.\end{tabular} &
  \begin{tabular}[c]{@{}l@{}}Broad application in readmission prediction modeling; \\ near-top performance in many applications.\end{tabular} \\ \bottomrule
\end{tabular}%
}
\end{table}

\subsection{Predictive Performance Results}\label{sec:sim_results}

\begin{figure}[!htbp]
    \centering
    \includegraphics[width=0.9\textwidth]{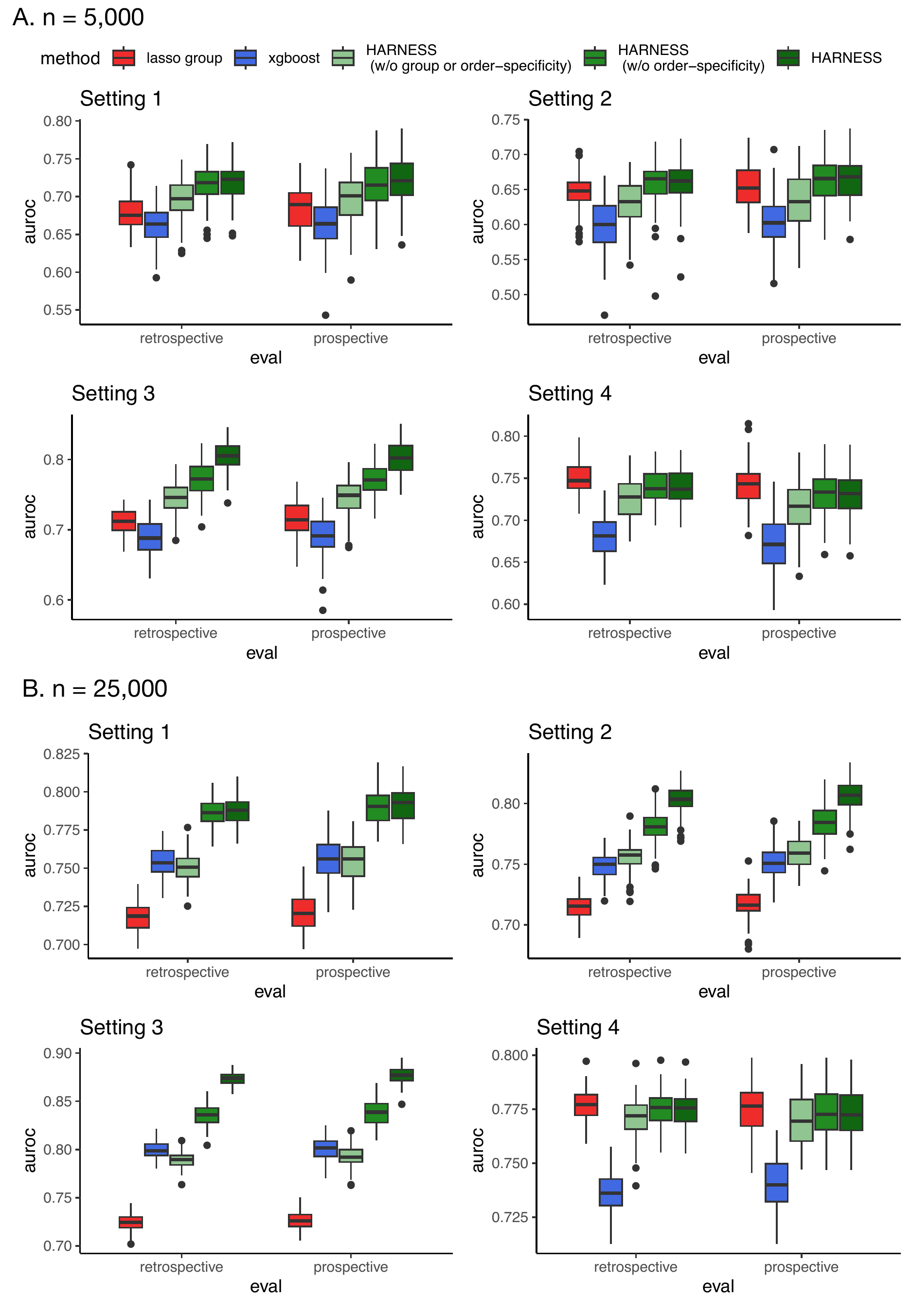}
    \caption{AUROC for retrospective and prospective performance evaluated under 4 different simulation settings for (A) small sample size ($n = 5,000$) and (B) large sample size ($n = 25,000$).}
    \label{fig:auroc_simulation}
\end{figure}

To assess both retrospective and prospective performance, 70\% of the data from the first nine years is randomly selected as the training set. The remaining 30\% of the data from the first nine years is used as testing set for retrospective performance; and the data from the last one year is used to evaluate  prospective performance, emulating the real world scenario where models are deployed and evaluated prospectively in time. While retrospective performance evaluation only illustrates performance on historical data and ignores the potential risk of overfitting, the additional evaluation of prospective performance of the models assesses their robustness and generalizability to future data, ensuring their practical utility for implementation in EHR data-driven applications. Model performance is assessed primarily based on the area under the curve (AUC) of the receiver operating characteristic (ROC). Figure \ref{fig:auroc_simulation} summarizes the AUROC (area under the receiver operating characteristic curve) for the retrospective and prospective performance of 100 replicates for different methods considered. The results measured in AUC for the precision-recall curve (PRAUC) are included in Section 3 of the supplementary materials for a more complete and accurate assessment of the predictive performance, since AUROC can be misleadingly high even with many false positives for imbalanced outcomes. 

\begin{table}[]
\caption{Simulation settings considered for performance evaluation of HARNESS and comparison methods.}
\label{tab:sim_setting}
\resizebox{\columnwidth}{!}{%
\begin{tabular}{@{}llll@{}}
\toprule
Setting &
  Name &
  Description &
  \begin{tabular}[c]{@{}l@{}}Expected best-performing\\ competing method\end{tabular} \\ \midrule
1 &
  \begin{tabular}[c]{@{}l@{}}Low second-order sparsity, \\ nonlinear-interaction-enriched design\end{tabular} &
  \begin{tabular}[c]{@{}l@{}}15 pairs of interactions involving the first 6 variables are included. \\ Specifically, $S = \{(1, 2), (1, 3), (1, 4), (1, 5), (1, 6), (2, 3), (2, 4)$,\\ $(2, 5), (2, 6),(3, 4), (3, 5), (3, 6), (4, 5), (4, 6), (5, 6)\}$.  \\ Each part of Equation \eqref{eq:simulation} has equal contribution in terms of variance.\end{tabular} &
  XGBoost \\ \midrule
2 &
  \begin{tabular}[c]{@{}l@{}}Medium second-order sparsity,\\ nonlinear-interaction-enriched design\end{tabular} &
  \begin{tabular}[c]{@{}l@{}}15 pairs of interactions involving the first 8 variables are included. \\ Specifically, $S = \{(1, 2), (1, 3), (1, 4), (1, 5), (1, 6), (1, 7), (1, 8)$,\\ $(1, 9),(1, 10), (2, 3), (2, 4), (2, 5), (2, 6), (2, 7), (2, 8)\}$.  \\ Each part of Equation \eqref{eq:simulation} has equal contribution in terms of variance.\end{tabular} &
  XGBoost \\ \midrule
3 &
  \begin{tabular}[c]{@{}l@{}}High second-order sparsity,\\ nonlinear-interaction-enriched design\end{tabular} &
  \begin{tabular}[c]{@{}l@{}}15 pairs of interactions involving the first 16 variables are included. \\ Specifically, $S = \{(1,2), (1,3), (1,4), (1,5), (1,6), (1,7), (1, 8), (1, 9)$,\\ $(1, 10),(1, 11), (1, 12), (1, 13), (1, 14), (1, 15), (1, 16)\}$.  \\ Each part of Equation \eqref{eq:simulation} has equal contribution in terms of variance.\end{tabular} &
  XGBoost \\ \midrule
4 &
  Linear-main-effects-only design &
  \begin{tabular}[c]{@{}l@{}}Only the linear one-way effects (the first part of Equation \eqref{eq:simulation}) \\ are included.\end{tabular} &
  lasso group \\ \bottomrule
\end{tabular}%
}
\end{table}

Figure \ref{fig:auroc_simulation} illustrates the retrospective and prospective performance (by AUROC) of all methods of interest under four simulation settings for sample size $n = 5,000$ (Figure \ref{fig:auroc_simulation} (A)) and $n = 25,000$ (Figure \ref{fig:auroc_simulation} (B)), respectively. While XGBoost is expected to be the best-performing comparison method for settings 1--3 due its ability to capture the underlying nonlinearity and interaction components, this pattern is observed only for the larger sample size $n = 25,000$. The underperformance to all other methods for $n = 5,000$ is likely due to its susceptibility to overfitting for relatively small sample size. For setting 4 with simple linear one-way effects only, XGBoost struggles to fit to the more simple underlying structure and remains the lowest-performing method for both sample sizes, as anticipated.

Although lasso group's robustness against overfitting overcomes its inability to capture the more complex underlying structure, achieving middle performance for smaller sample size $n=5,000$ in setting 1--3, it is eventually outperformed by all other methods for the larger sample size $n = 25,000$. For setting 4, lasso group remains the top-performing method for two different sample sizes as expected.

In terms of the performance of HARNESS, for settings 1--3 in sample size $n=5,000$, even with potentially compromised performance due to greater tendency to overfitting for smaller sample size $n = 5,000$, HARNESS still demonstrates some overperformance to the comparison methods, which is much more apparent for the larger sample size $n = 25,000$. For setting 4, HARNESS demonstrates its ability to adapt to simpler underlying structure and achieves equal performance with lasso group, which is most well-suited for this particular setting, demonstrating its ability to shrink effectively towards simpler modeling structures when the underlying data generating mechanism is simple and smooth.

For the two previous versions of HARNESS, first note that for each newer version, the performance is no worse than the previous one across all settings for both sample sizes, demonstrating the relative utility and benefit of incorporating each component of the new methodology. Specifically, improvements from HARNESS (w/o order-specificity) to HARNESS (w/o group or order-specificity) are observed for both sample sizes under all simulation settings, illustrating the greater ability of HARNESS (w/o order-specificity) to capture the simulated variability in coefficients across the groups. The improvement from HARNESS (w/o order-specificity) to HARNESS varies for different settings: improvement for setting 3 is the most apparent, followed by setting 2; and almost equal performance for these two versions is observed for setting 1 and 4. This difference in performance improvement is because the allowance of variable importance to be order-specific, the new part from HARNESS to HARNESS (w/o order-specificity), is incorporated to better handle the case where the two-way effects are not simply the products of the one-way effects. For setting 4 no two-way effect is included; and for setting 1 the two-way effects agree exactly the products of the first 6 variables having linear main effects in terms of coefficients and thus the order-specific variable importance is not needed. On the other hand, in Setting 2, interaction pairs have coefficients that deviate from the products of their main effects, with even greater deviations in Setting 3. In these settings, the introduction of order-specific variable importance thus further improves performance.

\section{A Case Study: 30-day Readmission Prediction among Those with Hematologic Diseases}

\subsection{Data Description}

Unplanned readmission, unplanned emergency department visits, and death within a short time interval after discharge are all considered undesirable post-hospitalization outcomes. Accurate risk prediction  implemented just prior to discharge from a hospital encounter helps identify high-risk patients, enabling more targeted follow-up care to reduce preventable readmission. Additionally, interpretations from the risk-prediction model suggests factors most contributive to readmission, which helps with early identification of high-risk patients for timely adjustments and more intensive monitoring during  inpatient visits.

We analyze the EHR data from 18,096 acute encounters including inpatient  admissions  and  observations from  of the population of patients whose primary diagnosis was hematologic in nature from December 30th, 2010 to December 1st, 2023 within ten integrated Midwestern hospitals. 
We focus on the hematologic population due to its markedly elevated burden: patients with hematologic diseases experience a 30-day all-cause readmission rate 70\% higher than the general rate, and their readmission costs exceed those of the corresponding index admission by 43.2\%, compared to a 12.4\% increase in the general population \citep{hcup2025}. The application to this population demonstrates improved prospective performance relative to the comparison methods.
The composite outcome rate for the above-mentioned three undesired outcomes is 36.17\%. A total of 1,185 clinical and non-clinical predictors are included in the model. Hematologic encounters are further grouped into 4 groups (level-1 groups) based on the pathophysiology: Genetic, Malignancy, Nutrition, and None. Within each level-1 group, the encounters can be further grouped based on the CCSR categories (level-2 groups). A detailed description of all groups is illustrated in Table \ref{tab:hem_group}. More details on data preprocessing are included in Section 2 of the Supplementary Materials.

\subsection{Predictive Performance Results}\label{sec:hem_results}

To assess both retrospective and prospective performance, data from the first 13 years (December 30th, 2010 to December 31st, 2022) are randomly split into training set (13,155 encounters from 8,626 patients) to train the model and testing set (3,393 encounters from 2,243 patients) for retrospective performance evaluation. The data from Jan 1st, 2023 to Dec 1st, 2023 (1,548 encounters from 1,225 patients), which is held out during training, is used as ``future" data for prospective performance assessment. 

\begin{figure}[htbp]
    \centering
    \includegraphics[width=1\textwidth]{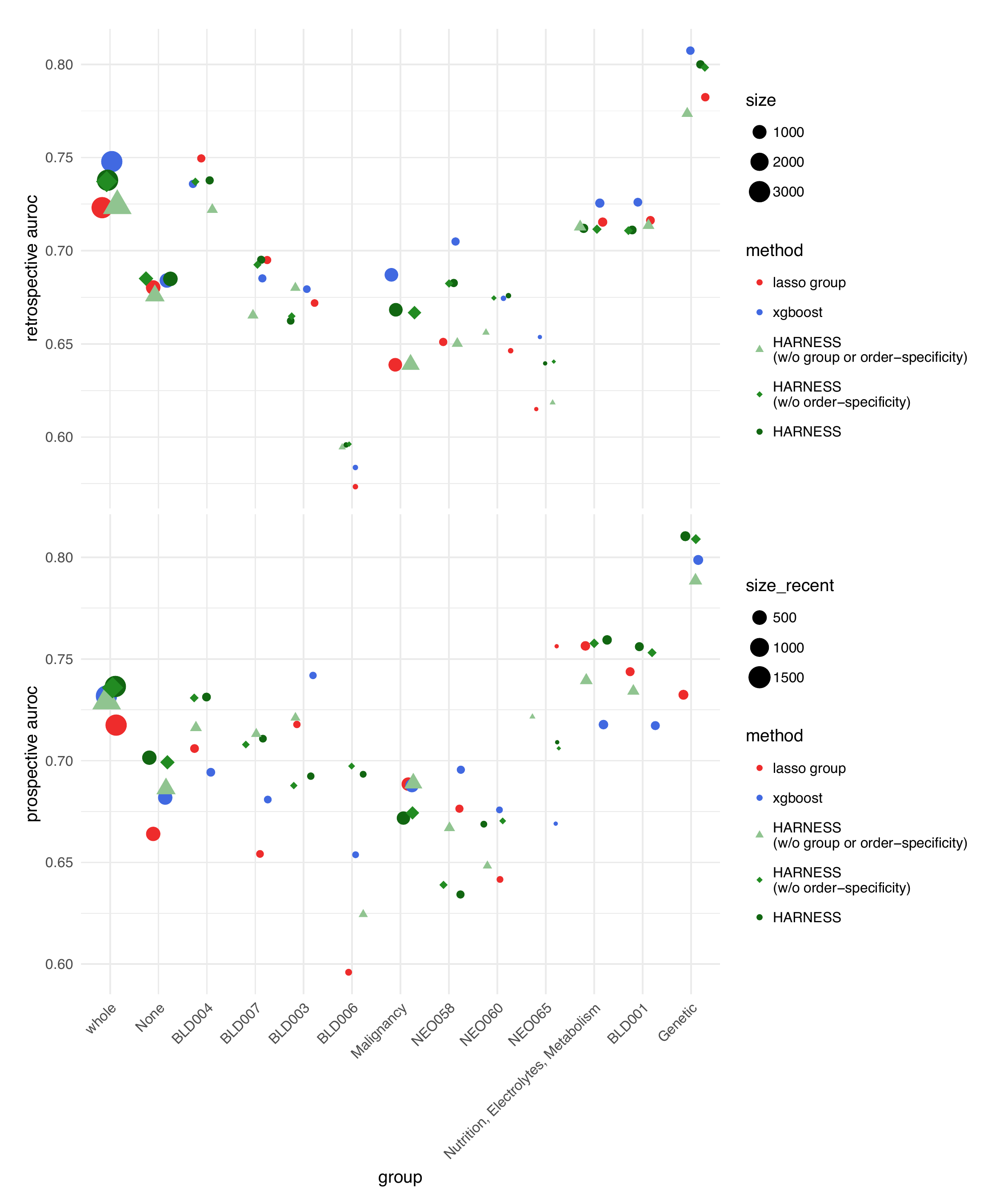}
    \caption{AUROC for retrospective (top) and prospective (bottom) performance evaluated at different pathophysiology (level-1) and CCSR (level-2) subgroups of the hematologic population. For prospective relevance, only subgroups with more than 100 samples in the most recent year (2023) are included.}
    \label{fig:auroc}
\end{figure}

For the retrospective performance measured in AUROC, as illustrated in the top panel of Figure \ref{fig:auroc}, XGBoost is the top-performing method for the whole hematologic population and most of the pathophysiology (level-1) subgroups and CCSR (level-2) subgroups. The only two exceptions are CCSR subgroups BLD004 and BLD007. The lasso group shows the worst performance for the whole hematologic population, and for the subgroup performance it is either the worst or in the middle most of the time, with BLD004 and BLD007 as only two exceptions. For HARNESS and the two simpler versions (HARNESS (w/o group or order-specificity) and HARNESS (w/o order-specificity)), each successive version results in performance no worse than the preceding version for the hematologic population and all subgroups, with BLD003 and BLD001 as only two exceptions. This illustrates the practical utility of the incorporation of group heterogeneity and order-specific variable importance to the model. However, even with the improvements from its simpler versions, HARNESS still performs in the middle in most  cases, and even the worst for the pathophysiology group Nutrition, Electrocytes, Metabolosim group and CCSR group BLD001. 

For the prospective performance measured in AUROC, as illustrated in the bottom panel of Figure \ref{fig:auroc}, XGBoost now outperforms only in BLD003 and NEO058, and performs in the middle and even the worst for other subgroups. This drop in performance is most likely due to the overfitting of XGBoost to the old data. Lasso group, albeit with less risk of overfitting and more capability of generalization to new data, is still not the top-performing method for the whole population or most of the subgroups, which is likely due to its over-simplified structural assumptions underfitting the data. As for HARNESS, first notice the similar trend of gradual improvement of performance from previous versions for all groups, with BLD003, Malignancy, and NEO058 as the only three exceptions, reinforcing the practical utility of the newly introduced methodology. In addition, with appropriate structural complexity balancing between fitting and generalization, HARNESS is the top performing method for the whole hematologic population and most of the subgroups. The performance measured in PRAUC is also included in Section 4 of supplementary materials.

\subsection{Visualization of Estimated Group Heterogeneity}

\begin{figure}[H]
    \centering
    \includegraphics[width=1\textwidth]{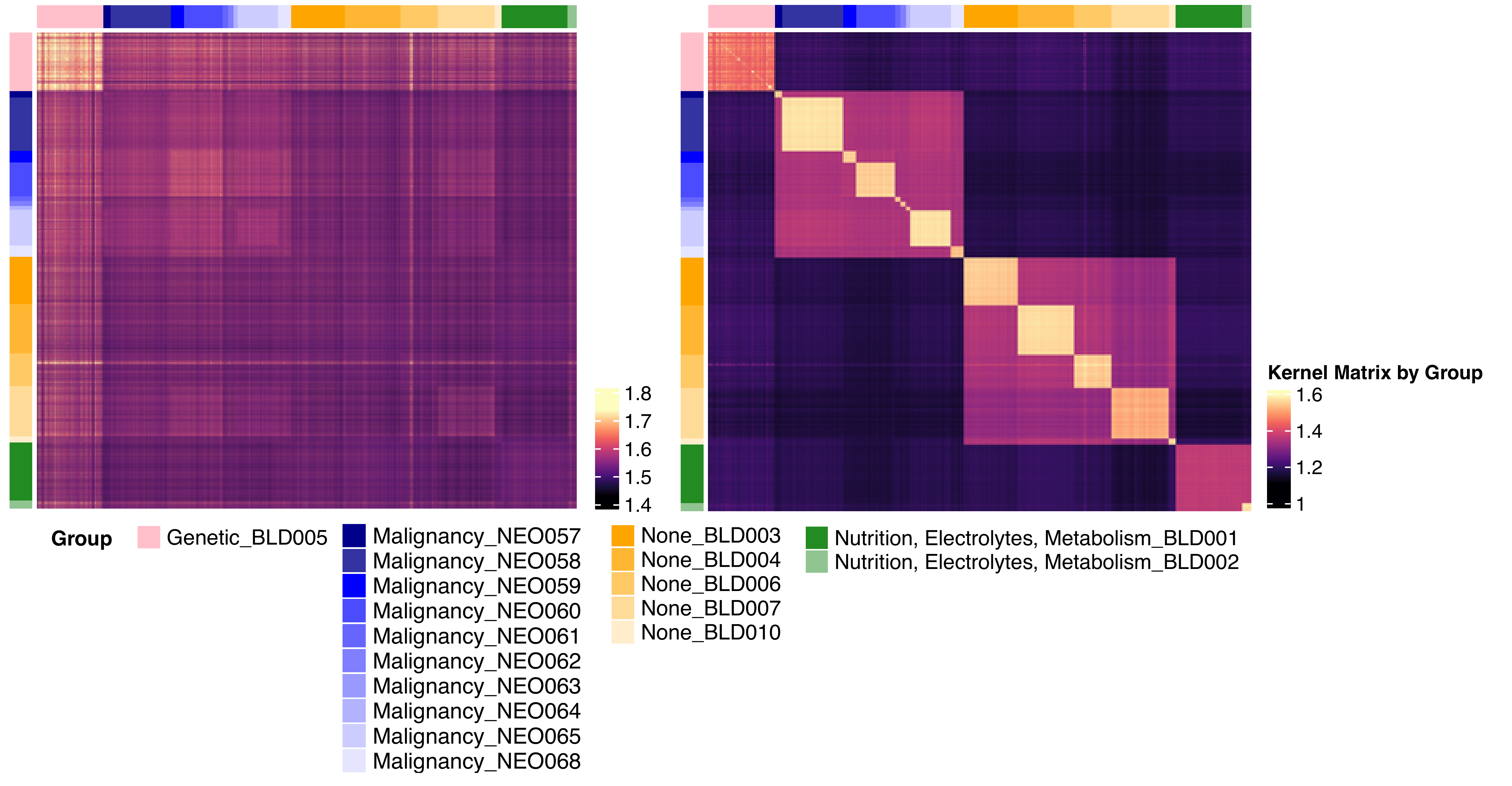}
    \caption{Visualization of the kernel matrix without the incorporation of grouping components (left) and with the grouping components (right).}
    \label{fig:kernel}
\end{figure}

Figure \ref{fig:kernel} visualizes the $K_{\text{HARNESS (w/o group or order-specificity)}}$ and $K_{\text{HARNESS}}$ for the whole Hematologic population hierarchically ordered by the subgroups. With the incorporation of group heterogeneity through group-specific kernels, encounters from the same pathophysiology (level-1) groups are encouraged to have greater similarity, which is further emphasized for same CCSR (level-2) groups. As for the comparison, $K_{\text{HARNESS (w/o group or order-specificity)}}$ shows group structure to some extent but not as apparent as the $K_{\text{HARNESS}}$.

\subsection{Interpretations of Order-specific Variable Importance}

\begin{figure}[htbp]
    \centering
    \includegraphics[width=1\textwidth]{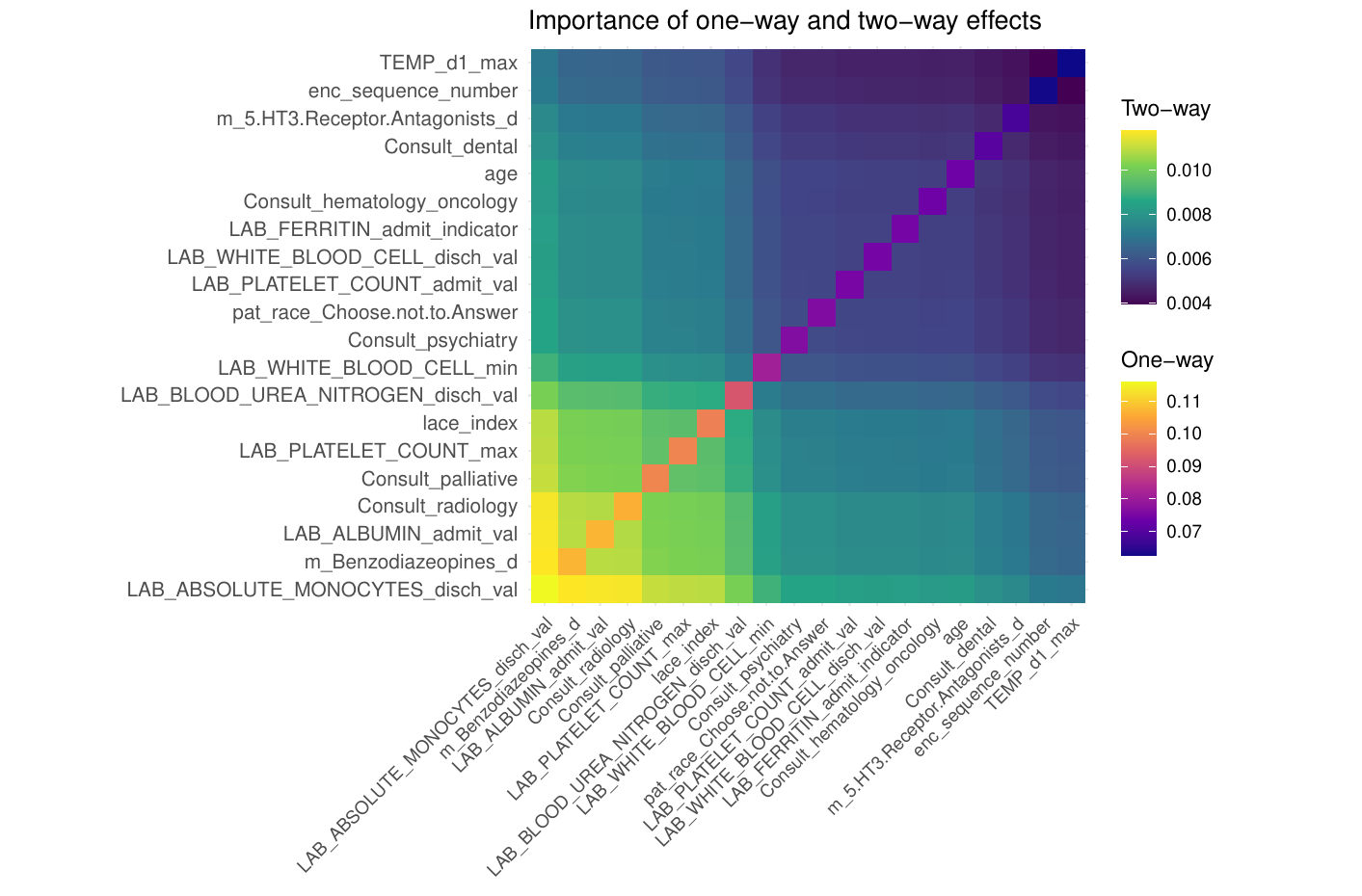}
    \caption{Variable importance in terms of one-way effects (diagonal) and two-way effects (off-diagonal) for top-20 most importance variables.}
    \label{fig:Vimp}
\end{figure}

Figure \ref{fig:Vimp} visualizes the variable importance in terms of the one-way effects and two-way effects for the top 20 most important variables identified as most predictive of readmission in the hematologic patient population. The majority consist of laboratory values and specialty consultations related to hematologic conditions, aligning with established clinical knowledge. Notably, the model also reveals less anticipated predictors—such as the patient’s decision to disclose racial identity—which may suggest novel avenues for early risk stratification and inpatient monitoring. The importance for two-way effects is generally proportional to the one-way importance of the variables involved.

\section{Discussion}

Accurate readmission prediction is  essential in helping prevent avoidable negative patient outcomes in health system settings. In this work, we propose a flexible approach for risk prediction of 30 day readmission based on hierarchical-group structure kernels. Our approach is designed to enhance predictive accuracy in high-dimensional settings for high-risk individuals while offering group-specific, clinically interpretable insights into key readmission drivers to support targeted interventions.

Given the complexity and heterogeneity in the inherent relationship between readmission and high-dimensional EHR features, we propose a new, multifaceted kernel which we call HARNESS to capture group-specific nonlinear and higher-order effects via a group-wise functional ANOVA decomposition. Specifically, HARNESS conducts order-specific variable selection through sparsity-inducing kernel summation, which varies across hierarchical clinical groups defined by subject-matter expertise based on CCSR categories to account for population heterogeneity. The kernel allows either models that fully vary by group or allows shrinkage towards a global model for all groups, enabling the borrowing of strength across related groups. Moreover, the kernel incorporates interpretable, order-specific variable importance measures, allowing for greater flexibility.

We apply HARNESS to EHR data arising from 18,096 acute encounters of those with hematologic diagnoses during 2010-2023 within ten integrated Midwestern hospitals. HARNESS outperforms competing methods, XGBoost and group-wise design-augmented lasso, in terms of prospective performance as measured by AUROC, both in the overall hematologic population and across the majority of pathophysiology and CCSR groups. The prominence of hematology-related variables among the top predictors reinforces the clinical relevance and face validity of the model. The emergence of less conventional factors suggests potential new contributors to readmission risk, meriting further exploration.

There are several remaining limitations that present opportunities for future methodological refinements. While our approach more accurately capture higher-order effects, even greater flexibility could be achieved through more advanced modifications, such as incorporating factorization machines to enhance the modeling of complex interactions. Additionally, our method enforces consistency in relative variable importance across groups, reflecting the overall contribution of features aggregated across all subgroups. However, enabling greater variation in variable importance could allow for more targeted subgroup-specific interpretations at the cost of additional kernel parameters. Furthermore, although the structural constraints in our model contribute to robustness in performance on unseen data, further improvements that explicitly aim to encourage robustness to variations in data or population may be important for the implementation of risk models in practice.

  \bibliography{HARNESS_0613}

\end{document}